\documentclass{article}

\usepackage[preprint]{neurips_2026}

\usepackage[utf8]{inputenc} 
\usepackage[T1]{fontenc}    
\usepackage{hyperref}       
\usepackage{url}            
\usepackage{booktabs}       
\usepackage{amsfonts}       
\usepackage{nicefrac}       
\usepackage{microtype}      
\usepackage{xcolor}         

\usepackage{subcaption}
\usepackage{amsmath} 
\usepackage{algorithm}
\usepackage{algpseudocode}
\usepackage{graphicx} 
\usepackage{wrapfig}
\usepackage{booktabs}
\usepackage{longtable}
\usepackage{array}
\usepackage{float}

\newcommand{\AlgInput}{\Statex \textbf{Input: }}
\newcommand{\AlgOutput}{\Statex \textbf{Output: }}

\usepackage{booktabs}
\usepackage{xcolor,colortbl}
\definecolor{best}{RGB}{200,245,200}      
\definecolor{second}{RGB}{255,250,200}    
\definecolor{third}{RGB}{255,230,200}     
\definecolor{worst}{RGB}{255,200,200}     

\title{

HeLoCo: Efficient asynchronous low-communication training under data and device heterogeneity}

\author{%
\begin{tabular}{c}
\textbf{Abdullah Al Asif}\textsuperscript{1}
\quad
Patrick Diem\textsuperscript{2}
\quad
\textbf{Juan Pablo Mu\~noz}\textsuperscript{3} \\
\textbf{Felix Wolf}\textsuperscript{2}
\quad
\textbf{Ali Jannesari}\textsuperscript{1}
\quad
\textbf{Arya Mazaheri}\textsuperscript{2,4} \\[1.5mm]
\end{tabular}
\\
\textsuperscript{1}Iowa State University \\
\textsuperscript{2}Technical University of Darmstadt \\
\textsuperscript{3}Maro Systems \\
\textsuperscript{4}PanocularAI \\[1mm]
}

\begin{document}

\maketitle

\begin{abstract}
Distributed Low-Communication (DiLoCo) training reduces communication overhead by allowing workers to perform multiple local optimization steps before sending pseudo-gradients to a global outer update. Its asynchronous variant further improves hardware utilization by removing synchronization barriers, but at the cost of stale pseudo-gradients computed from outdated model states. As a result, these updates can become misaligned with the current global optimization direction, particularly in heterogeneous systems. This issue becomes even more pronounced when data are non-IID, a setting that has not been well studied in asynchronous low-communication training. To address this limitation, we propose \textbf{HeLoCo}, a direction-aware correction method for asynchronous low-communication training that uses outer momentum as a reference for the current optimization trajectory and selectively adjusts incoming pseudo-gradients before the outer update. Updates that remain aligned are preserved, while directionally conflicting components are corrected. On multilingual language-model training with heterogeneous workers and non-IID data, HeLoCo consistently improves validation loss. It outperforms existing asynchronous DiLoCo-based baselines by up to 7.5\% at a fixed token budget, exceeds asynchronous momentum look-ahead by up to 3.3\% at a fixed wall-clock budget, and surpasses the synchronous baseline by up to 22.1\% under severe system heterogeneity.
Our analysis further shows how staleness, worker speed, and data heterogeneity shape update quality and convergence in highly decentralized and heterogeneous training setups.

\end{abstract}

\section{Introduction}
\label{sec:introduction}

Training large language models (LLMs) typically relies on tightly coupled, high-performance clusters with high-bandwidth, low-latency interconnects. While effective within a single data center, this setup becomes challenging when compute resources are geographically distributed or heterogeneous. In such cases, communication slows down and synchronization can take up a lot of time \citep{Douillardetal2023,douillard2025streaming}. When multiple workers train together, some may end up waiting for others due to slow connections or imbalances in workload. To solve this issue, the Distributed Low Communication (DiLoCo) algorithm lets worker nodes perform several updates locally before sending their updates, called pseudo-gradients~\citep{Douillardetal2023}. By decoupling local and global updates, DiLoCo reduces communication requirements, while preserving training efficiency, making it particularly well suited to loosely connected clusters.\citep{Douillardetal2023,jaghouar2024opendiloco,douillard2025streaming}.

\begin{wrapfigure}{r}{0.5\textwidth}
    \centering
    \includegraphics[width=0.48\textwidth]{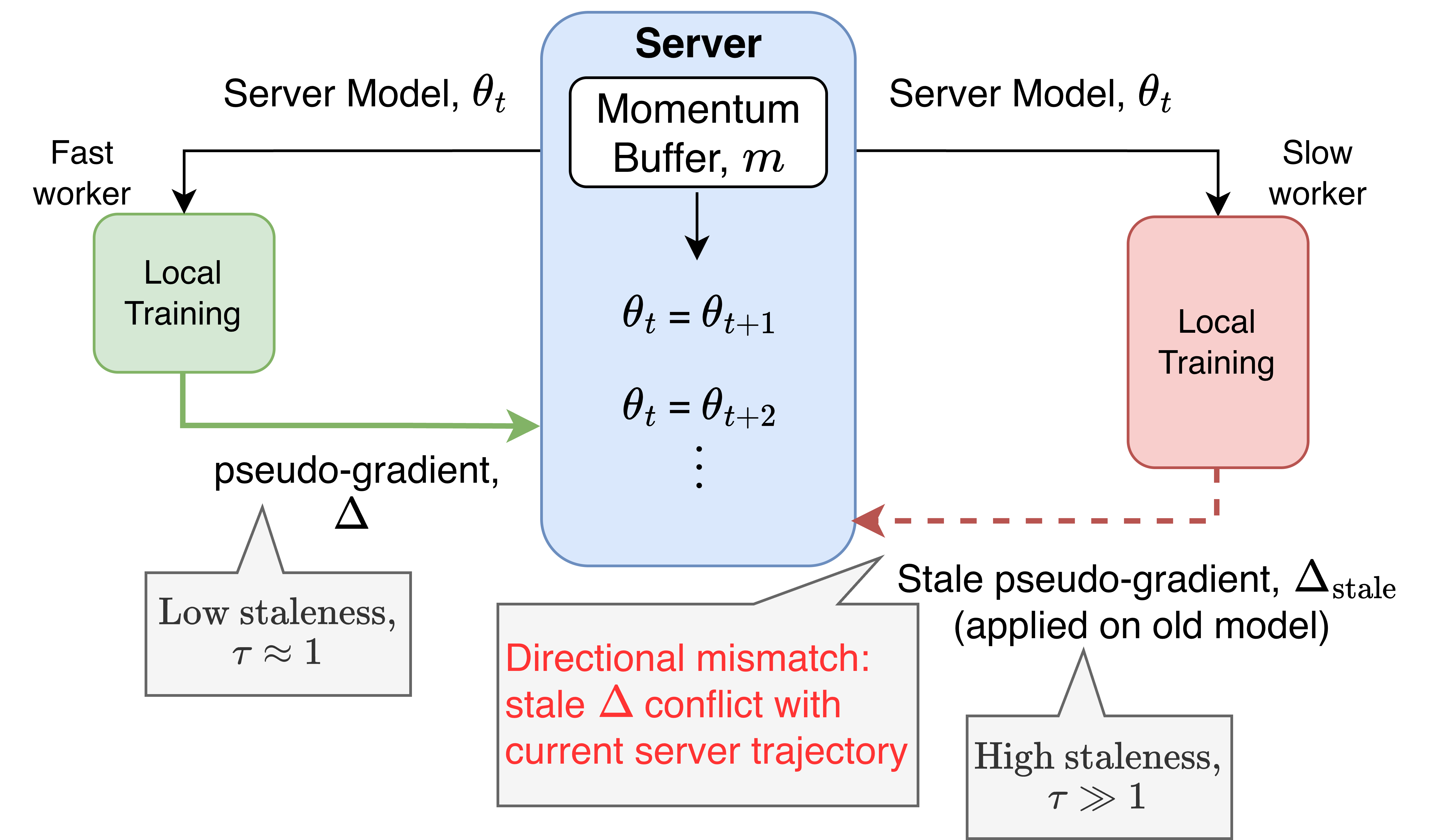}
    \caption{Staleness asymmetry in asynchronous DiLoCo with heterogeneous workers.}
    \label{fig:staleness}
\end{wrapfigure}

Synchronous DiLoCo requires all workers to send their updates before starting the next global step. When workers operate at different speeds or are distributed across regions, these differences can introduce stragglers, leading to delays and idle time. Switching to asynchronous variants allows each worker's updates to be applied immediately. This improves both hardware efficiency and overall performance. However, a key challenge in this setting is gradient staleness, meaning that some gradient estimates are based on outdated model versions, which introduces inconsistency when they are consumed by the outer optimizer. Prior work~\citep{liu2024asynchronous} shows that this inconsistency can degrade convergence. As a result, although the asynchronous variant of DiLoCo permits more frequent updates than its synchronous counterpart, it can potentially yield worse optimization performance. Figure~\ref{fig:staleness} illustrates the gap in synchronization between fast and slow workers. Fast workers contribute updates based on the most recent global model, ensuring high relevance. In contrast, slow workers (often called "stragglers") submit gradients derived from outdated model parameters. Because the global optimizer has already progressed, these "stale" updates can be counterproductive or even lead to training instability.


Existing correction methods only partially address the problem. Traditional delayed-update methods apply an exponential discount to updates based on their age or delay compensation \citep{zinkevich2009slow,zheng2017asynchronous}. Recent asynchronous DiLoCo schemes introduce corrections using delayed Nesterov-like and momentum lookahead techniques \citep{liu2024asynchronous,ajanthan2025momentum}.  However, these methods only apply corrections to the entire pseudo-gradient, which proves to be too coarse for heterogeneous systems and non-IID (Independent and Identically Distributed) dataset distributions. An update from a slower worker is not only more delayed but may also be less compatible with the current distribution utilized by the global trajectory \citep{li2020federated}. These two effects do not necessarily impact all parameter blocks of the pseudo-gradient equally; some may align with the current direction while others may contradict it, potentially hindering the convergence of the outer method.

To address this issue, we propose \textbf{HeLoCo} (\textbf{He}terogeneity-aware \textbf{Lo}w-\textbf{Co}mmunication training), a direction-aware correction algorithm designed for asynchronous low-communication training. The core idea is to use outer momentum as a reference for determining the trustworthiness of each parameter block, where each block is an individual model tensor, such as a weight matrix. This approach allows processing of various stale gradient directions in alignment with the current outer update process. We further include a lightweight momentum-based~\citep{zhang2019lookahead} initialization step to improve directional consistency.


Our contributions are as follows:
\begin{itemize}
    \item We identify the directional mismatch as a key failure mode in asynchronous DiLoCo: under system heterogeneity and non-IID data, individual tensor blocks within a stale pseudo-gradient can conflict with the outer update direction.

    \item We propose HeLoCo, a momentum-guided correction method that initializes workers with a look-ahead model and selectively keeps or adjusts the pseudo-gradient before the outer update.

    \item Empirically, our method reduces validation loss by up to 7.5\% over asynchronous Nesterov and 3.3\% over asynchronous MLA at a fixed token budget, and by 22.1\% over synchronous Nesterov under both system and data heterogeneity.
\end{itemize}

\section{Background and Related Work}
\label{sec:related_work}

Stale pseudo-gradients in asynchronous distributed training have been studied 
from the perspective of delay compensation and momentum correction, but the 
question of whether different parts of the same update deserve different 
treatment has received little attention. We review the three lines of work 
most relevant to this gap.

\paragraph{Low-communication training for language models.}
DiLoCo showed that separating local SGD steps from a global outer optimizer dramatically cuts synchronization frequency without sacrificing model quality~\citep{Douillardetal2023}. OpenDiLoCo confirmed this recipe holds even across continents~\citep{jaghouar2024opendiloco}. Follow-up work pushed further: Streaming DiLoCo overlaps communication with computation to reduce peak bandwidth~\citep{douillard2025streaming}, DiLoCoX scales to larger models via parallelism and compression~\citep{qi2025dilocox}, and NoLoCo~\citep{kolehmainen2025noloco} explores dropping global synchronization altogether. These works share a common focus on \emph{when} and \emph{how often} workers communicate. We ask a different question: given that some updates will inevitably arrive late, can we do better than applying them as-is?

\paragraph{Asynchrony and stale-update correction.}
Stale gradients are as old as asynchronous optimization itself. Early work showed they can be tolerated in sparse settings~\citep{recht2011hogwild}, and parameter-server systems exposed their broader effects~\citep{zinkevich2009slow, dean2012large}. A subtler insight came later: asynchronous execution does not just add noise, it implicitly introduces a momentum-like bias~\citep{mitliagkas2016asynchrony}. This makes the outer momentum a natural reference for diagnosing staleness. Taylor-style compensation was one early attempt to exploit this~\citep{zheng2017asynchronous}. More recently, MLA~\citep{ajanthan2025momentum} corrects staleness by extrapolating the negative momentum direction uniformly across the entire pseudo-gradient. We argue, however, that tensor blocks within the same stale update can have fundamentally different alignment with the outer trajectory, and a single uniform correction is too coarse to handle this.

\paragraph{Heterogeneity, non-IID data, and directional conflict.}
The staleness problem is further compounded when workers operate at different speeds and train on non-IID data — conditions that are unavoidable in globally distributed training. In such heterogeneity, stale updates do not just arrive late; they carry gradients shaped by a different data distribution and a different point in the loss landscape~\citep{mcmahan2017fedavg}. This directional distortion has been well documented in federated learning, where proximal regularization~\citep{li2020federated}, control variates~\citep{karimireddy2020scaffold}, and normalized averaging~\citep{wang2020fednova} were each designed to counter it. In multi-task optimization, PCGrad made a related observation: gradient conflicts are better resolved geometrically, block by block, than through a single global correction~\citep{yu2020pcgrad}. Our method brings this geometric intuition into asynchronous DiLoCo, using outer momentum as a reference to selectively preserve, dampen, or reorient each tensor block of a stale update — depending on how well it agrees with the direction the outer optimizer is already moving.

\section{Methodology}
\label{sec:method}


Section~\ref{sec:introduction} highlights the shortcomings of asynchronous training with low communication. The local update might be based on an old version of the global model, leading to issues in scenarios where there are heterogeneous worker update speeds and non-IID data. Despite this, delayed pseudo-gradients preserve some of the local gradient information; however, some directions in the tensor may not correspond with the latest global model update directions. In order to solve this issue, HeLoCo employs two strategies: one for initialization, allowing workers to start with a look-ahead model state, and another for alignment, ensuring returned gradients properly align with the current update direction.



\paragraph{Problem setup.}
We consider distributed training over $K$ workers. Worker $i$ samples data from
a local distribution $\mathcal{D}_i$, and the target objective is:
\begin{equation}
    \min_{\theta} F(\theta)
    =
    \sum_{i=1}^{K} p_i F_i(\theta),
    \qquad
    F_i(\theta)=\mathbb{E}_{z\sim\mathcal{D}_i}
    [\ell(\theta;z)],
    \label{eq:global_objective}
\end{equation}
where $\theta$ denotes the model parameters, $p_i\geq0$,
$\sum_i p_i=1$, and $\ell(\theta;z)$ is the training loss. The weights $p_i$ define the intended contribution of each local objective to the global objective. The local distributions $\mathcal{D}_i$ may be non-IID, and workers may have different computation or communication speeds. By contrast, in asynchronous training, update arrivals are driven by worker speed, so faster workers can influence the outer model more frequently than slower workers.


At outer step $t$, the global model maintains parameters $\theta_t$ and an outer
momentum buffer $m_t$. Here, \emph{outer} refers to the global DiLoCo-level optimization step performed when a worker returns a pseudo-gradient, while \emph{inner} refers to the local optimization steps performed independently on each worker. When worker $i$ becomes available at outer step $s_i$, the synchronizer
initializes it with an outer model state $\bar{\theta}_{s_i}$. The worker
initializes its local model from this state and performs $H$ inner optimization
steps:
\begin{equation}
    \theta_{i,0}^{(s_i)}=\bar{\theta}_{s_i},
    \qquad
    \theta_{i,h+1}^{(s_i)}
    =
    \theta_{i,h}^{(s_i)}
    -
    \alpha_h g_i(\theta_{i,h}^{(s_i)}),
    \quad h=0,\ldots,H-1,
    \label{eq:local_update}
\end{equation}
where $\alpha_h$ is the inner learning rate and $g_i(\cdot)$ is a stochastic
gradient on worker $i$'s local data. After local training, the worker returns
the pseudo-gradient:
\begin{equation}
    \Delta_i^{(s_i)}
    =
    \bar{\theta}_{s_i}
    -
    \theta_{i,H}^{(s_i)}.
    \label{eq:pseudo_grad}
\end{equation}
This sign convention makes $\Delta_i^{(s_i)}$ the descent displacement
accumulated by local training, so the synchronizer applies it through a negative
outer step.

If the update arrives at outer step $t$, its staleness is defined as:
\begin{equation}
    \tau_i(t)=t-s_i.
    \label{eq:staleness}
\end{equation}
Thus, $\Delta_i^{(s_i)}$ was computed from the initial model
$\bar{\theta}_{s_i}$, while the outer model now holds $\theta_t$. This delay is the
source of the stale-update error that HeLoCo is designed to reduce.

\paragraph{Momentum-guided worker initialization.}
The first use of outer momentum occurs before local training begins.
Instead of initializing a worker with the current outer model $\theta_r$ at
outer step $r$, HeLoCo initializes it with a look-ahead model:
\begin{equation}
    \bar{\theta}_r
    =
    \theta_r-\eta_r\mu m_r,
    \label{eq:dispatch}
\end{equation}
where $\eta_r$ is the outer learning rate at the time of initialization and
$\mu$ is the outer momentum coefficient. This model is a one-step prediction of
where the outer model is moving under its current momentum. Since the worker will
return after some delay, starting local training from $\bar{\theta}_r$ reduces the positional gap between the worker's initial model and the outer model state at arrival.

For a worker initialized at step $s_i$, the returned pseudo-gradient is
therefore:
\begin{equation}
    \Delta_i^{(s_i)}
    =
    \left(\theta_{s_i}-\eta_{s_i}\mu m_{s_i}\right)
    -
    \theta_{i,H}^{(s_i)}.
    \label{eq:pseudo_grad_lookahead}
\end{equation}
This look-ahead initialization reduces the mismatch at the beginning of local
training. However, it cannot guarantee that the returned pseudo-gradient will remain directionally useful after several outer updates. This motivates a second decision when the pseudo-gradient returns: which tensor blocks should be kept unchanged, and which should be corrected before the outer update?

\paragraph{Measuring update alignment.}
When $\Delta_i^{(s_i)}$ arrives at outer step $t$, HeLoCo compares each tensor
block of the update with the corresponding tensor block of the current outer models'
momentum $m_t$. For readability, we write $\Delta_b$ for a block of the
arriving pseudo-gradient and $(m_t)_b$ for the corresponding momentum block.
All norms and inner products are computed after vectorizing each tensor block;
$\|\cdot\|$ denotes the Euclidean norm, equivalently the Frobenius norm for
tensor-shaped parameters.

If both blocks have non-negligible norm, define:
\begin{equation}
    \hat{u}_b
    =
    \frac{\Delta_b}{\|\Delta_b\|},
    \qquad
    \hat{v}_b
    =
    \frac{(m_t)_b}{\|(m_t)_b\|}.
    \label{eq:unit_vectors}
\end{equation}
The directional compatibility score is defined as:
\begin{equation}
    c_b=\hat{u}_b^\top\hat{v}_b.
    \label{eq:cosine}
\end{equation}
A positive value indicates that the stale tensor still agrees with the current global model trajectory. A negative value indicates that the tensor contains a component opposing the momentum direction. A value near zero indicates that the tensor is not directly conflicting but is only weakly supported by the current trajectory.

This tensor-wise view is the key design choice in HeLoCo. A stale
pseudo-gradient is not treated as uniformly good or uniformly harmful. Within
the same worker update, some tensors may remain aligned with the current global models'
trajectory, while others may conflict with it. A single scalar penalty on the
entire pseudo-gradient cannot make this distinction.

\paragraph{Correcting stale directions.}
HeLoCo applies a geometric correction to each tensor block. If either
$\|\Delta_b\|$ or $\|(m_t)_b\|$ is below the numerical threshold
$\varepsilon$, the momentum direction is unreliable for that block and the
block is passed through unchanged. Let $c_{\mathrm{ok}}$ denote the threshold for sufficient alignment. If the block is sufficiently aligned, $c_b\geq c_{\mathrm{ok}}$, it is also preserved:
\begin{equation}
    \widehat{\Delta}_b=\Delta_b.
    \label{eq:aligned_pass}
\end{equation}
This leaves unchanged the tensor blocks that already agree with the current
outer update direction.

If $c_b<0$, the block contains an anti-momentum component. HeLoCo reduces only
that conflicting component rather than discarding the entire tensor:
\begin{equation}
    \widehat{\Delta}_b
    =
    \Delta_b
    -
    \beta_b c_b\|\Delta_b\|\hat{v}_b,
    \label{eq:project}
\end{equation}
where
\begin{equation}
    \beta_b
    =
    \min\left\{
    k_s(-c_b)\mathrm{conf}_b,\,
    \beta_{\max}
    \right\}.
    \label{eq:beta}
\end{equation}
Here $k_s>0$ controls the shrinkage strength and $\beta_{\max}\in(0,1]$ prevents overcorrection. Since $c_b<0$, the correction reduces the component that points against the current outer update direction. Geometrically, this can be interpreted as a partial projection-based correction along the normalized momentum direction $\hat{v}_b$. The incoming block $\Delta_b$ can be decomposed into a parallel component $(\Delta_b^\top \hat{v}_b)\hat{v}_b$ and an orthogonal residual $\Delta_b-(\Delta_b^\top \hat{v}_b)\hat{v}_b$. HeLoCo attenuates only the anti-momentum part of the parallel component, while leaving the orthogonal residual unchanged so that local information not directly conflicting with the current trajectory is retained.


For weakly aligned blocks, $0\leq c_b<c_{\mathrm{ok}}$, HeLoCo does not remove a component. Instead, it gently reorients the block toward the current momentum direction while preserving its magnitude:
\begin{align}
    \tilde{u}_b
    &=
    (1-\lambda_b)\hat{u}_b+\lambda_b\hat{v}_b,
    \label{eq:reorient_interp}
    \\
    \widehat{\Delta}_b
    &=
    \|\Delta_b\|
    \frac{\tilde{u}_b}{\max\{\|\tilde{u}_b\|,\varepsilon\}},
    \label{eq:reorient_scale}
\end{align}
with
\begin{equation}
    \lambda_b
    =
    \min\left\{
    k_d(1-c_b)\mathrm{conf}_b,\,
    1
    \right\}.
    \label{eq:lambda}
\end{equation}
The coefficient $k_d>0$ controls the reorientation strength. The less aligned the block is, the more strongly it is pulled toward the momentum direction. In the non-degenerate case where $\|\tilde{u}_b\|\geq\varepsilon$, this operation
preserves $\|\widehat{\Delta}_b\|=\|\Delta_b\|$ and changes only the direction of the tensor.

The confidence factor used in both correction cases is defined as:
\begin{equation}
    \mathrm{conf}_b
    =
    \frac{\|\Delta_b\|}
    {\|\Delta_b\|+\kappa\|(m_t)_b\|+\varepsilon},
    \label{eq:conf}
\end{equation}
where $\kappa>0$ controls the relative scale of the momentum norm. This factor makes the correction conservative when the arriving block is small relative to the accumulated momentum and stronger when the arriving block is large.

After all blocks are processed, the corrected pseudo-gradient is reassembled as:
\begin{equation}
    \widehat{\Delta}
    =
    \{\widehat{\Delta}_b\}_b.
    \label{eq:corrected_delta}
\end{equation}
Appendix~\ref{app:heloco_theory} provides a theoretical justification
for the tensor-wise correction used by HeLoCo.

\paragraph{Outer-model update.}
The outer update applies the corrected pseudo-gradient immediately. If the asynchronous training rule uses a worker or delay weight, which we denote by $\rho_t$; otherwise, $\rho_t=1$.
The weighted corrected update is:
\begin{equation}
    G_t=\rho_t\widehat{\Delta}.
    \label{eq:weighted_delta}
\end{equation}
The global model then updates its momentum and parameters as
\begin{align}
    m_{t+1}
    &=
    \mu m_t+(1-\mu)G_t,
    \label{eq:mom_update}
    \\
    \theta_{t+1}
    &=
    \theta_t-\eta_t\left(G_t+\mu m_{t+1}\right).
    \label{eq:theta_update}
\end{align}
The updated momentum $m_{t+1}$ is used in the parameter step, allowing the
arriving pseudo-gradient to first update the outer direction estimate and then contribute to the model update. 



The HeLoCo communication method works the same way as DiLoCo. Each worker processes a single model, updates it locally, and creates a single pseudo-gradient vector. However, the way they prepare for data exchange is different. Before local training starts, the synchronizer sets up the look-ahead model. After this, to prevent incorrect updates, the synchronizer modifies the gradients using momentum before applying the outer update procedure. This adjustment only needs to access the existing momentum buffer and requires an extra pass through the gradient vector, adding a cost of $O(d)$ per arrival. Appendix~\ref{app:notation} summarizes the notation used in this section, and Appendix~\ref{app:heloco_algorithm} provides the HeLoCo pseudocode.

\section{Experiments and Analyses}
\label{sec:results}


\paragraph{Setup.} We evaluate HeLoCo using TinyGPT~\citep{radford2019language} on the multilingual
mC4~\citep{Raffel2020} dataset. Across experiments, we compare Nesterov, MLA, and HeLoCo under different worker-speed and data heterogeneity settings. Appendix~\ref{app:repro_compute} provides implementation,
hardware, and runtime details.

\subsection{Convergence Under Heterogeneity}
\label{subsec:convergence_results}

\begin{figure*}[t]
    \centering
    \includegraphics[width=\textwidth]{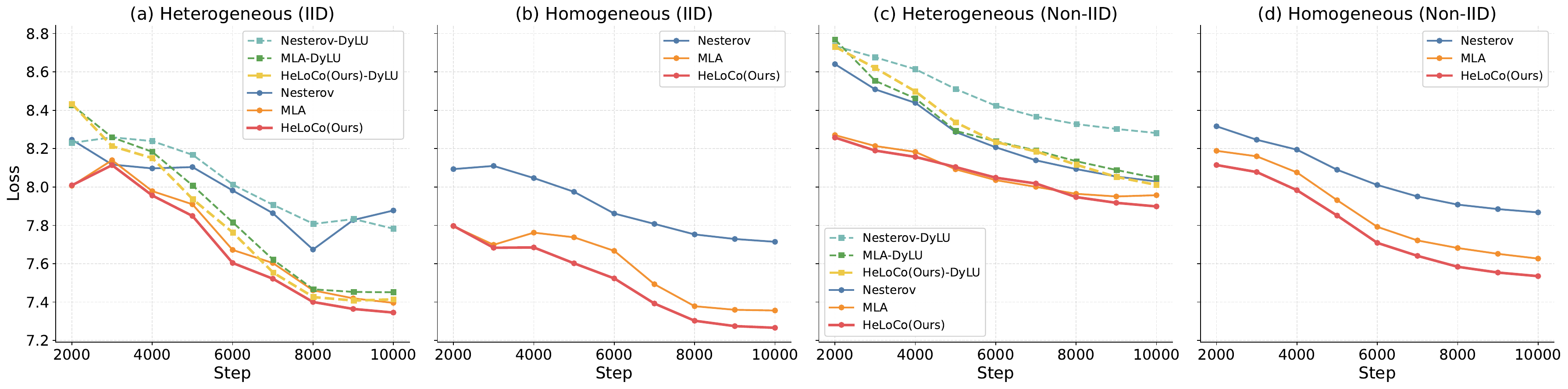}
    \caption{Validation loss over training steps on C4 under IID and non-IID
    data distributions with heterogeneous and homogeneous worker speeds. DyLU
    variants are shown only for heterogeneous settings.}
    \label{fig:iid_noniid_15m}
\end{figure*}

For the convergence experiments, we compare methods based on two key factors: how fast the workers are (either all the same speed or different speeds) and how the data is organized (either similar or different types). In the setting with different speeds, five workers take between 0.74 and 7.50 seconds for each step. This leads to an uneven update process: the fastest worker provides 58.2\% of all updates and has an average delay of 0.72. Meanwhile, the two slowest workers together account for only 12\% of the updates, with delays of 14.38 and 16.50, respectively. We also look at Dynamic Local Updates (DyLU)~\citep{wang2019adaptive}, which helps balance how often workers provide updates by changing the number of local steps they take. With DyLU, contributions from workers become almost equal (19-22\%), and the average staleness is approximately $\bar{\tau}\approx 4$. Crucially, DyLU is orthogonal to HeLoCo, as DyLU balances worker participation by changing the local-update schedule, while HeLoCo corrects returned pseudo-gradients before the outer update.
Figure~\ref{fig:iid_noniid_15m} compares training loss across IID and non-IID data settings under both heterogeneous and homogeneous worker configurations. When we use IID data on heterogeneous device~\ref{fig:iid_noniid_15m}(a), HeLoCo outperforms the other methods after about 6000 steps, achieving a final loss of 7.35. In comparison, MLA has a final loss of 7.41, and Nesterov has 7.89. Since the data are IID, this improvement mainly comes from the time delays introduced by workers with different speeds. In the homogeneous IID setting~\ref{fig:iid_noniid_15m}(b), the difference between HeLoCo and MLA decreases, as expected. When workers move at similar speeds, and delays are minimal, most parts of the data align well with the recent global model updates and remain unchanged during corrections. 

The non-IID results highlight the importance of correcting finer-grained components when both statistical heterogeneity and system heterogeneity are present. In the heterogeneous non-IID setting, HeLoCo shows a clear advantage, with a final loss of 7.91, compared to 7.96 for MLA and 8.29 for Nesterov. In this scenario, outdated gradients often mix different levels of alignment, so correcting at the global update level is not precise enough. In the homogeneous, non-IID setting, our method again outperforms MLA and Nesterov. This shows that having non-IID data alone can cause enough differences to make per-block corrections beneficial, even with low delays. The DyLU variants also provide a helpful comparison. DyLU lowers staleness variance by equalizing how often updates occur, but it also reduces local training time for some workers. In both IID and non-IID settings, DyLU variants do not consistently perform better than the corresponding non-DyLU HeLoCo runs. This suggests that, in our case, keeping local optimization strong and fixing stale updates at the outer optimizer is more effective than cutting local training time to reduce staleness.


\paragraph{Non-IID data domain analysis.}

\begin{figure*}[t]
    \centering
    \includegraphics[width=0.98\textwidth]{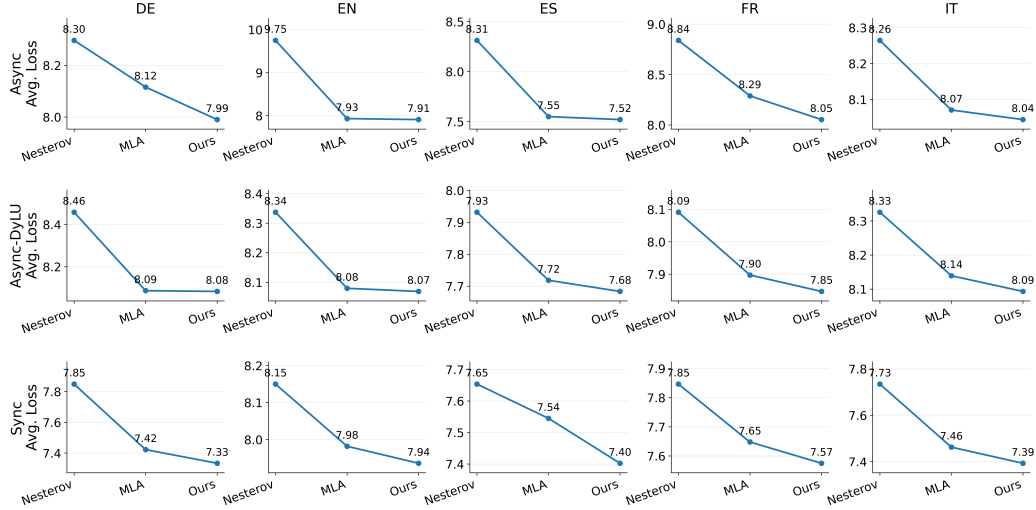}
    \caption{Evaluation loss comparison across five languages (de, en, es,
    fr, it) on multilingual C4 under non-IID training with five workers.}
    \label{fig:loss_multilingual_comparison}
\end{figure*}


Figure~\ref{fig:loss_multilingual_comparison} shows the results for different languages, where each language represents a distinct data domain with its own distribution, and how HeLoCo improves with staleness. Each language has a fixed-speed worker, keeping staleness the same during training. English dataset, trained by the most active worker (58.2\% of training updates, average staleness 0.72), shows only a slight gain over MLA (7.91 vs. 7.93). This happens because most parameter blocks are aligned with the global model and need little correction. German and French, handled by slower workers (average staleness of 14.38 and 5.12, respectively), show larger improvements of 0.13 and 0.24, respectively. In these cases, stale gradients cause misalignment, and HeLoCo's correction focuses only on the misaligned parts, preserving useful information rather than discarding everything. The Async-DyLU row confirms this: when DyLU forces all workers to similar
staleness ($\bar{\tau}\approx 4$), our approach still outperforms MLA, although by a smaller margin. This shows that per-block correction helps even when the staleness is the same. The synchronous row serves as a staleness-free baseline, where HeLoCo still outperforms MLA due to differences in the data. The slowest workers (contributing only 6.4\% and 5.6\% of total training updates) show fewer benefits from correction. This is expected, since infrequent updates mean the real issue shifts from misalignment to insufficient updates. These findings suggest that this method works best under moderate staleness: enough misalignment to benefit from correction, but enough updates for those corrections to matter.

\subsection{Effect of worker staleness in non-IID setting}
\label{sec:noniidinsights}
Building on Section~\ref{subsec:convergence_results}, which evaluates a single fixed worker pace in the heterogeneous setting, this section examines how varying worker pace configurations affects optimization and loss dynamics under non-IID data distributions. Specifically, we analyze cases~c) and~d) of Figure~\ref{fig:iid_noniid_15m} in greater depth to demonstrate HeLoCo's robustness across a spectrum of heterogeneous worker configurations. We compare HeLoCo against three baselines: (i) synchronous DiLoCo with Nesterov, (ii) asynchronous DiLoCo with Nesterov, and (iii) asynchronous DiLoCo with MLA. Following the findings of Section~\ref{subsec:convergence_results}, we exclude DyLu variants as they yield no performance gain in any configuration, and use synchronous Nesterov as the primary reference to assess HeLoCo and MLA in terms of loss progression over both steps and wall-clock time.

Each worker is assigned a speed drawn from $\{1, 2, 6, 15\}$ (seconds per step). We construct scenarios with progressively increasing staleness, culminating in the extreme configuration $(1,15,15,15,15)$, in which one fast worker and four highly stale workers operate concurrently. Training runs for 300 outer steps of 80 inner steps each (24k total steps). Hyperparameters are listed in Appendix~\ref{app:hyperparameters} and match those from \citep{liu2024asynchronous}.

\begin{figure}
    \centering
    \includegraphics[width=1\linewidth]{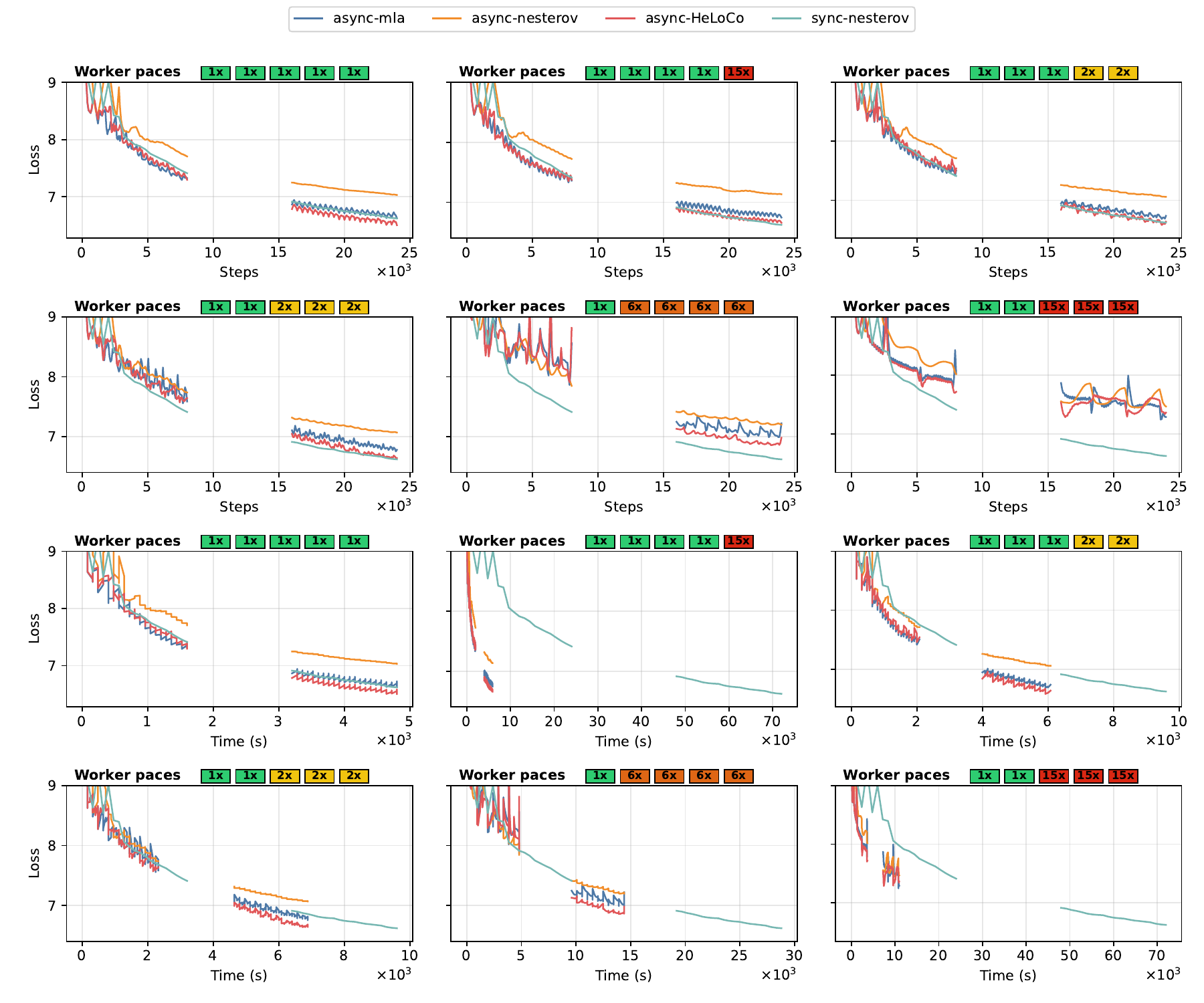}
    \caption{Validation loss over steps (left) and wall-clock time (right) for different worker pace configurations in the non-IID setting. Worker-pace colors indicate staleness level: green for low staleness, yellow for moderate staleness, and red for high staleness.}
    \label{fig:workerplot}
    \vspace{-1em}
\end{figure}

Figure~\ref{fig:workerplot} and Table~\ref{tab:fixedlosstable} (left) show training loss over a fixed token budget across staleness configurations. HeLoCo consistently achieves lower loss than MLA in most settings. In the homogeneous case $(1,1,1,1,1)$ and several other low-staleness configurations, HeLoCo also outperforms synchronous Nesterov, indicating that tensor-wise correction is beneficial even in the absence of staleness. As staleness grows, the performance gap between methods widens. Asynchronous Nesterov exhibits unstable behavior and diverges across all configurations. In high-staleness regimes, synchronous Nesterov surpasses both MLA and asynchronous Nesterov in token-budget loss; however, HeLoCo remains consistently more robust, achieving lower loss than MLA across nearly all configurations. The maximum token-budget improvement of HeLoCo over MLA is 3.3\%. In extreme cases such as $(1,15,15,15,15)$, synchronous Nesterov achieves substantially lower loss than both asynchronous methods at a fixed token budget.

This picture changes when runtime is taken into account. Due to synchronization barriers, the wall-clock time of synchronous training is bottlenecked by the slowest worker. As a result, asynchronous variants exhibit an advantage in loss-vs.-time performance across most configurations, as shown in Figure~\ref{fig:workerplot} and Table~\ref{tab:fixedlosstable} (right). HeLoCo achieves the best trade-off between convergence speed and loss in all but two configurations; in the extreme staleness setting $(1,15,15,15,15)$, it falls short of the synchronous baseline. At its best, HeLoCo's time-budget loss is 22.1\% lower than that of synchronous Nesterov. While HeLoCo typically improves over MLA, the margin is configuration-dependent: in moderate-staleness regimes the gains are consistent, whereas in the most extreme staleness cases a slight degradation is observed, suggesting that adaptive mechanisms for managing highly stale updates remain an open direction.

\begin{table}[h]
\centering
\caption{Validation loss for a fixed token budget (left) and a fixed time budget (right) across worker pace configurations. Left: cell color indicates the best (green) and worst (red) method per row. Right: color indicates HeLoCo's improvement (green) or degradation (red) relative to each baseline at matched wall-clock time. \emph{Legend:} L-HeLoCo: HeLoCo loss; L-AMLA: async MLA loss; L-AN: async Nesterov loss; L-SN: sync Nesterov loss; $\Delta X$: relative improvement of HeLoCo over $X$ at 24k steps; T$\Delta X$: relative improvement of HeLoCo over $X$ at time $T$.}
\tiny
\label{tab:fixedlosstable}
\begin{tabular}{l rrrrrrr|rrrrr}
\toprule
Worker pace & L-HeLoCo & L-AMLA & L-AN & L-SN & $\Delta$AMLA & $\Delta$AN & $\Delta$SN & T & T$\Delta$AMLA & T$\Delta$AN & T$\Delta$SN \\
\midrule
 &  &  &  & & (\%) & (\%) & (\%) & (s) & (\%) & (\%) & (\%) \\
 \midrule

(1,6,6,6,6) &
\cellcolor{second}6.98 &
\cellcolor{worst}7.22 &
\cellcolor{third}7.20 &
\cellcolor{best}6.62 &
3.33 & 3.04 & -5.43 & 14400 & \cellcolor{best}3.33 & \cellcolor{best}3.04 & \cellcolor{best}5.80 \\

(1,2,2,2,2) &
\cellcolor{best}6.54 &
\cellcolor{third}6.72 &
\cellcolor{worst}7.04 &
\cellcolor{second}6.62 &
2.62 & 7.08 & 1.22 & 8000 & \cellcolor{best}2.62 & \cellcolor{best}7.08 & \cellcolor{best}3.04 \\

(1,1,6,6,6) &
\cellcolor{second}6.69 &
\cellcolor{third}6.85 &
\cellcolor{worst}7.29 &
\cellcolor{best}6.62 &
2.45 & 8.23 & -5.48 & 9600 & \cellcolor{best}0.92 & \cellcolor{best}3.05 & \cellcolor{best}4.68 \\

(1,1,1,6,6) &
\cellcolor{second}6.66 &
\cellcolor{third}6.80 &
\cellcolor{worst}7.15 &
\cellcolor{best}6.62 &
2.03 & 6.75 & -0.62 & 7200 & \cellcolor{best}2.03 & \cellcolor{best}6.75 & \cellcolor{best}13.04 \\

(1,1,2,2,2) &
\cellcolor{second}6.64 &
\cellcolor{third}6.78 &
\cellcolor{worst}7.07 &
\cellcolor{best}6.62 &
1.99 & 6.00 & -0.34 & 6880 & \cellcolor{best}1.99 & \cellcolor{best}6.00 & \cellcolor{best}3.13 \\

(1,1,1,1,1) &
\cellcolor{best}6.50 &
\cellcolor{third}6.63 &
\cellcolor{worst}7.03 &
\cellcolor{second}6.62 &
1.84 & 7.50 & 1.78 & 4800 & \cellcolor{best}1.84 & \cellcolor{best}7.50 & \cellcolor{best}1.78 \\

(1,15,15,15,15) &
\cellcolor{second}7.64 &
\cellcolor{third}7.78 &
\cellcolor{worst}7.81 &
\cellcolor{best}6.62 &
1.75 & 2.15 & -15.39 & 19200 & \cellcolor{best}1.75 & \cellcolor{best}2.15 & \cellcolor{worst}-0.26 \\

(1,1,1,2,2) &
\cellcolor{second}6.64 &
\cellcolor{third}6.74 &
\cellcolor{worst}7.06 &
\cellcolor{best}6.62 &
1.48 & 5.98 & -0.22 & 6080 & \cellcolor{best}1.48 & \cellcolor{best}5.98 & \cellcolor{best}10.46 \\

(1,1,1,1,6) &
\cellcolor{best}6.60 &
\cellcolor{third}6.69 &
\cellcolor{worst}7.10 &
\cellcolor{second}6.62 &
1.36 & 7.10 & 0.40 & 5760 & \cellcolor{best}1.36 & \cellcolor{best}7.10 & \cellcolor{best}15.78 \\

(1,1,1,1,15) &
\cellcolor{second}6.66 &
\cellcolor{third}6.75 &
\cellcolor{worst}7.13 &
\cellcolor{best}6.62 &
1.30 & 6.67 & -0.54 & 5920 & \cellcolor{best}1.30 & \cellcolor{best}6.67 & \cellcolor{best}22.07 \\

(1,1,1,1,2) &
\cellcolor{best}6.61 &
\cellcolor{third}6.69 &
\cellcolor{worst}7.05 &
\cellcolor{second}6.62 &
1.23 & 6.22 & 0.16 & 5360 & \cellcolor{best}1.23 & \cellcolor{best}6.22 & \cellcolor{best}10.80 \\

(1,1,1,15,15) &
\cellcolor{second}6.84 &
\cellcolor{third}6.89 &
\cellcolor{worst}7.21 &
\cellcolor{best}6.62 &
0.75 & 5.22 & -3.22 & 7680 & \cellcolor{best}0.75 & \cellcolor{best}5.22 & \cellcolor{best}18.85 \\

(1,1,15,15,15) &
\cellcolor{third}7.36 &
\cellcolor{second}7.29 &
\cellcolor{worst}7.47 &
\cellcolor{best}6.62 &
-1.00 & 1.42 & -11.17 & 10960 & \cellcolor{worst}-1.00 & \cellcolor{best}1.42 & \cellcolor{best}7.81 \\

\bottomrule
\end{tabular}
\end{table}

\section{Conclusion}
\label{sec:conclusion}

We presented HeLoCo, a correction method for asynchronous low-communication training that targets the core failure mode of gradient staleness: directional mismatch between stale pseudo-gradients and the current outer optimization trajectory. Rather than applying a single global correction, HeLoCo operates at the tensor-block level, using outer momentum as a trust reference to selectively preserve, attenuate, or interpolate each arriving update. The result is a method that remains aligned with the outer optimizer under both system and data heterogeneity, without discarding gradient information unnecessarily.
Our experiments confirm that staleness is not a uniform problem: the benefit of correction scales with per-worker staleness and is most pronounced when system heterogeneity and non-IID data compound. In this regime, HeLoCo consistently outperforms asynchronous baselines at both fixed token and wall-clock time budgets, and often matches or exceeds synchronous training without its synchronization overhead. These results suggest a broader principle for asynchronous low-communicatoin distributed training, in which reducing communication frequency is not enough and the direction of each arriving update must also be actively managed. 

\paragraph{Limitations and future work.}
We conducted the evaluation using five workers and 15M-parameter models on subsets of C4 dataset. This controlled setting allows us to study HeLoCo under varying heterogeneous conditions, but it does not yet capture the scale of practical low-communication LLM training. Future work includes testing HeLoCo with more workers, larger models, longer training periods, and more diverse datasets. An interesting next step is to combine HeLoCo with decoupled DiLoCo-style systems~\citep{douillard2026decoupled} to explore larger and more diverse deployments.  Moreover, our results indicate that extreme staleness is still challenging asynchronous settings. As mentioned in Appendix~\ref{app:deepiid}, highly stale pseudo-gradients may become unreliable, which suggests that discarding them may sometimes be preferable to correcting them. One promising idea is to merge HeLoCo with adaptive staleness-aware filtering, where we adjust, down-weight, or discard a returned pseudo-gradient based on how stale it is and its alignment statistics.

\subsection*{Acknowledgements}
We gratefully acknowledge the support from Germany’s Federal Ministry of Breakthrough Innovation (SPRIN-D) through the Composite Learning Challenge under the SymphonyLearn project, and the hessian.AI Service Center (funded by the Federal Ministry of Research, Technology and Space, BMFTR, grant no. 16IS22091), the hessian.AI Innovation Lab (funded by the Hessian Ministry for Digital Strategy and Innovation, grant no. S-DIW04/0013/003). We would also like to thank the National Science Foundation (NSF) for supporting this project under grant 2211982 and 2426580. This work utilized the Delta system at the National Center for Supercomputing Applications (NCSA) through allocation CIS240626. We also acknowledge support from the National Science Foundation under grants 2138259, 2138286, 2138307, 2137603, and 2138296.






\bibliographystyle{plainnat}
\bibliography{ref} 

\clearpage
\appendix
\section{Appendix}

\subsection{Notation}
\label{app:notation}

Table~\ref{tab:notation} summarizes the notation used in this paper

{\small
\setlength{\LTleft}{0pt}
\setlength{\LTright}{0pt}
\renewcommand{\arraystretch}{1.12}

\begin{longtable}{p{0.18\linewidth} p{0.76\linewidth}}
\caption{Symbols used in the methodology and appendix algorithm.}
\label{tab:notation}\\

\toprule
\textbf{Symbol} & \textbf{Description} \\
\midrule
\endfirsthead

\toprule
\textbf{Symbol} & \textbf{Description} \\
\midrule
\endhead

\midrule
\multicolumn{2}{r}{\emph{Continued on next page}} \\
\endfoot

\bottomrule
\endlastfoot

\multicolumn{2}{l}{\textbf{Distributed training objective}}\\
\midrule
$K$ & Total number of workers participating in distributed training. \\
$i,j$ & Worker indices. Typically, $i$ denotes the worker whose update arrives, while $j$ denotes an available worker being initialized. \\
$\mathcal{D}_i$ & Local data distribution of worker $i$. These distributions may be non-IID across workers. \\
$F(\theta)$ & Global training objective. \\
$F_i(\theta)$ & Local objective of worker $i$, defined over its local data distribution. \\
$p_i$ & Nonnegative weight of worker $i$ in the global objective, where $\sum_{i=1}^{K} p_i = 1$. \\
$z$ & Training sample drawn from a local data distribution. \\
$\ell(\theta; z)$ & Loss of model parameters $\theta$ on sample $z$. \\
$\theta$ & Generic model parameter vector. \\
$d$ & Total number of scalar model parameters, used in the overhead discussion. \\

\midrule
\multicolumn{2}{l}{\textbf{Asynchronous training setup}}\\
\midrule
$t$ & Current server step at which a returned pseudo-gradient is processed. \\
$r$ & Server step at which an available worker is initialized. \\
$s_i$ & Initialization step of worker $i$, i.e., the server step when worker $i$ receives its initial model. \\
$\tau_i(t)$ & Staleness of worker $i$'s update at server step $t$, defined as $\tau_i(t) = t - s_i$. \\
$\theta_t$ & Server model at server step $t$. \\
$\theta_r$ & Server model at initialization step $r$. \\
$m_t$ & Server momentum buffer at step $t$, representing the current server trajectory. \\
$m_r$ & Server momentum buffer at initialization step $r$. \\
$\mu$ & Outer momentum coefficient used by the server. \\
$\eta_t$ & Outer learning rate at server step $t$. \\
$\eta_r$ & Outer learning rate used when constructing the look-ahead model at initialization step $r$. \\

\midrule
\multicolumn{2}{l}{\textbf{Local worker training}}\\
\midrule
$\bar{\theta}_{s_i}$ & Initial model sent to worker $i$ at step $s_i$. In HeLoCo, this is the look-ahead model. \\
$\bar{\theta}_r$ & Look-ahead model used to initialize an available worker, defined as $\bar{\theta}_r = \theta_r - \eta_r \mu m_r$. \\
$H$ & Number of local optimization steps performed by a worker before returning its pseudo-gradient. \\
$h$ & Index of a local optimization step, where $h = 0, \ldots, H-1$. \\
$\theta_{i,h}^{(s_i)}$ & Local model of worker $i$ after $h$ inner steps, initialized from server step $s_i$. \\
$\theta_{i,0}^{(s_i)}$ & Initial local model of worker $i$, equal to $\bar{\theta}_{s_i}$. \\
$\theta_{i,H}^{(s_i)}$ & Final local model of worker $i$ after $H$ local steps. \\
$\alpha_h$ & Inner learning rate used at local step $h$. \\
$g_i(\cdot)$ & Stochastic gradient computed on worker $i$'s local data. \\
$\Delta_i^{(s_i)}$ & Worker pseudo-gradient, defined as $\Delta_i^{(s_i)} = \bar{\theta}_{s_i} - \theta_{i,H}^{(s_i)}$. \\

\midrule
\multicolumn{2}{l}{\textbf{Tensor-wise compatibility and correction}}\\
\midrule
$b$ & Index of a tensor block, such as a weight tensor or bias tensor. \\
$\Delta_b$ & Tensor block $b$ of the arriving pseudo-gradient. \\
$(m_t)_b$ & Tensor block $b$ of the current server momentum. \\
$\|\cdot\|$ & Euclidean norm after vectorization; equivalent to the Frobenius norm for tensor-shaped parameters. \\
$\hat{u}_b$ & Unit direction of the pseudo-gradient block, defined as $\hat{u}_b = \Delta_b / \|\Delta_b\|$. \\
$\hat{v}_b$ & Unit direction of the momentum block, defined as $\hat{v}_b = (m_t)_b / \|(m_t)_b\|$. \\
$c_b$ & Directional compatibility score between the pseudo-gradient block and the momentum block, defined as $c_b = \hat{u}_b^\top \hat{v}_b$. \\
$c_{\mathrm{ok}}$ & Threshold above which a tensor block is considered sufficiently aligned and is kept unchanged. \\
$\varepsilon$ & Small numerical threshold used to avoid division by zero and to detect near-zero unreliable directions. \\
$\widehat{\Delta}_b$ & Corrected version of tensor block $\Delta_b$. \\
$\widehat{\Delta}$ & Full corrected pseudo-gradient after all tensor blocks are processed. \\
$\mathrm{conf}_b$ & Confidence factor controlling how strongly tensor block $b$ should be corrected. \\
$\kappa$ & Positive constant controlling how the momentum norm affects $\mathrm{conf}_b$. \\
$k_s$ & Positive coefficient controlling shrinkage strength when a tensor block conflicts with momentum. \\
$\beta_b$ & Tensor-wise shrinkage coefficient used when $c_b < 0$. \\
$\beta_{\max}$ & Upper bound on $\beta_b$ to prevent overcorrection. \\
$k_d$ & Positive coefficient controlling reorientation strength for weakly aligned tensor blocks. \\
$\lambda_b$ & Tensor-wise interpolation coefficient used when $0 \leq c_b < c_{\mathrm{ok}}$. \\
$\tilde{u}_b$ & Interpolated direction between the pseudo-gradient direction and the momentum direction. \\

\midrule
\multicolumn{2}{l}{\textbf{Outer model update}}\\
\midrule
$\rho_t$ & Optional arrival weight applied to the corrected pseudo-gradient. If unused, $\rho_t = 1$. \\
$G_t$ & Weighted corrected update, defined as $G_t = \rho_t \widehat{\Delta}$. \\
$m_{t+1}$ & Updated global momentum after incorporating the arriving corrected update. \\
$\theta_{t+1}$ & Updated global model after applying the outer update. \\

\midrule
\multicolumn{2}{l}{\textbf{Local variables in the pseudocode}}\\
\midrule
$u$ & Tensor block of the returned pseudo-gradient, $u=(\Delta_i^{(s_i)})_b$. \\
$v$ & Corresponding momentum block, $v=(m_t)_b$. \\
$\hat{u}$ & Normalized update direction, $\hat{u}=u/\|u\|$. \\
$\hat{v}$ & Normalized momentum direction, $\hat{v}=v/\|v\|$. \\
$\tilde{u}$ & Direction obtained by blending $\hat{u}$ with $\hat{v}$ for weakly aligned blocks. \\

\end{longtable}
}

\subsection{Theoretical Justification of the HeLoCo Correction}
\label{app:heloco_theory}

This appendix gives a local theoretical justification for the tensor-wise
correction used by HeLoCo. The result is not a convergence proof for the full
asynchronous training process. Instead, it shows that, for each non-degenerate
tensor block, the correction does not reduce the signed component along the
current outer momentum direction and does not increase the block norm.

The correction is motivated by the same geometry used in gradient surgery
methods such as PCGrad~\cite{yu2020pcgrad}. PCGrad treats two gradients as
conflicting when their cosine similarity is negative and removes the
conflicting projection component. HeLoCo uses this idea in a different setting:
it compares each stale pseudo-gradient block with the corresponding outer
momentum block. Unlike PCGrad, HeLoCo applies a damped correction in the
anti-aligned case and also adjusts weakly aligned blocks by blending their
direction with the momentum direction. For the proof only, we use shorthand notation; these symbols correspond to the notation in Table~\ref{tab:notation}.


\paragraph{Lemma.}
Fix a tensor block $b$. Let
\[
    u=\Delta_b,
    \qquad
    v=(m_t)_b
\]
denote the returned pseudo-gradient block and the corresponding outer momentum
block. Assume $\|u\|,\|v\|>\varepsilon$, $\beta_{\max}\leq1$, and
$\varepsilon<1/\sqrt{2}$. Define
\[
    \hat{u}=\frac{u}{\|u\|},
    \qquad
    \hat{v}=\frac{v}{\|v\|},
    \qquad
    c=\hat{u}^{\top}\hat{v}.
\]
Here $c$ is the cosine similarity between the returned block and the momentum
block. Let $\widehat{\Delta}_b$ be the corrected block produced by HeLoCo.
Then
\[
    \widehat{\Delta}_b^{\top}\hat{v}
    \geq
    u^{\top}\hat{v},
    \qquad
    \|\widehat{\Delta}_b\|\leq \|u\|.
\]
Thus, HeLoCo does not reduce the signed component of the block along the
outer momentum direction and does not amplify the returned block.

\paragraph{Proof.}
If $c\geq c_{\mathrm{ok}}$, HeLoCo leaves the block unchanged, so both claims
hold with equality.

If $c<0$, HeLoCo applies
\[
    \widehat{\Delta}_b
    =
    u-\beta_b c\|u\|\hat{v},
    \qquad 0\leq \beta_b\leq 1.
\]
Since $u^\top\hat{v}=c\|u\|$, we have
\[
    \widehat{\Delta}_b^\top\hat{v}
    =
    (1-\beta_b)u^\top\hat{v}
    \geq
    u^\top\hat{v},
\]
because $u^\top\hat{v}<0$. Also,
\[
    \|\widehat{\Delta}_b\|^2
    =
    \|u\|^2\left(1-\beta_b(2-\beta_b)c^2\right)
    \leq
    \|u\|^2.
\]
Therefore, counter-directed blocks become less opposed to the momentum direction
without norm amplification.

If $0\leq c<c_{\mathrm{ok}}$, HeLoCo forms
\[
    \tilde{u}=(1-\lambda_b)\hat{u}+\lambda_b\hat{v},
    \qquad 0\leq\lambda_b\leq1.
\]
Since $0\leq c\leq1$, we have
\[
    \|\tilde{u}\|^2
    =
    1-2\lambda_b(1-\lambda_b)(1-c)
    \geq
    \frac{1}{2}.
\]
Because $\varepsilon<1/\sqrt{2}$, the numerical guard is inactive in this
case, and
\[
    \widehat{\Delta}_b
    =
    \|u\|\frac{\tilde{u}}{\|\tilde{u}\|}.
\]
Thus $\|\widehat{\Delta}_b\|=\|u\|$. Let
\[
    c'
    =
    \frac{\tilde{u}^{\top}\hat{v}}{\|\tilde{u}\|}
\]
be the cosine similarity after correction. Then
\[
    c'
    =
    \frac{(1-\lambda_b)c+\lambda_b}
    {\sqrt{(1-\lambda_b)^2+\lambda_b^2
    +2\lambda_b(1-\lambda_b)c}} .
\]
For $0\leq c\leq1$ and $0\leq\lambda_b\leq1$, the numerator is at least $c$
and the denominator is at most $1$, so $c'\geq c$. Therefore,
\[
    \widehat{\Delta}_b^\top\hat{v}
    =
    \|u\|c'
    \geq
    \|u\|c
    =
    u^\top\hat{v}.
\]
Hence weakly aligned blocks are moved toward the outer momentum direction
without changing their norm.

The lemma applies to non-degenerate blocks for which the cosine comparison is
well defined. When $\|u\|<\varepsilon$ or $\|v\|<\varepsilon$, HeLoCo skips the
correction because the update or momentum direction is numerically unreliable.
This completes the proof.

\paragraph{Interpretation.}
The lemma shows that HeLoCo does not blindly suppress stale pseudo-gradients.
Well-aligned blocks are preserved, anti-aligned blocks have their opposing
component damped, and weakly aligned blocks are adjusted toward the momentum
direction without increasing their magnitude. Since outer momentum summarizes
recent outer updates, this local guarantee supports the use of momentum as the
reference direction for correcting stale pseudo-gradient blocks.

\subsection{HeLoCo Algorithm}
\label{app:heloco_algorithm}
\begin{algorithm}[H]
\caption{Asynchronous DiLoCo with HeLoCo}
\label{alg:main}
\begin{algorithmic}[1]
\AlgInput
\begin{tabular}[t]{@{}l@{}}
Initial model $\theta^{(0)}$, outer momentum $m_0{=}0$, outer learning rate $\eta$,\\
momentum coefficient $\mu$, $k$ workers, data shards $\{\mathcal{D}_1,\ldots,\mathcal{D}_k\}$
\end{tabular}
\AlgOutput
\begin{tabular}[t]{@{}l@{}}
Updated global model $\theta_{t+1}$ and outer momentum $m_{t+1}$
\end{tabular}
\For{outer step $t = 1, 2, \ldots$}
    \If{worker $j$ is available at step $r$}
        \State $\bar{\theta}_r \gets \theta_r - \eta_r \mu m_r$
        \Comment{\textbf{HeLoCo}: look-ahead initialization}
        \State Send $\bar{\theta}_r$ to worker $j$; record $s_j \gets r$
    \EndIf
    \For{inner step $h = 1 \ldots H$}
        \State $x \sim \mathcal{D}_j$
        \State $\mathcal{L} \gets f(x, \theta_j^{(t)})$
        \State $\theta_j^{(t)} \gets \texttt{InnerOpt}(\theta_j^{(t)}, \nabla_\mathcal{L})$
    \EndFor
    \If{worker $i$ returns $\Delta_i^{(s_i)}$ at step $t$}
        \State $\tau_i \gets t - s_i$
        \Comment{Measure staleness}
        \State $\widehat{G}_t \gets \textsc{BlockCorrect}(\Delta_i^{(s_i)}, m_t, \rho_t)$
        \Comment{\textbf{HeLoCo}: per-block correction (Algorithm~\ref{alg:correct})}
        \State $m_{t+1} \gets \mu m_t + (1-\mu)\widehat{G}_t$
        \State $\theta_{t+1} \gets \theta_t - \eta\bigl(\widehat{G}_t + \mu m_{t+1}\bigr)$
    \EndIf
\EndFor
\end{algorithmic}
\end{algorithm}

\begin{algorithm}[H]
\caption{Tensor Directional Correction}
\label{alg:correct}
\begin{algorithmic}[1]
\AlgInput
\begin{tabular}[t]{@{}l@{}}
Pseudo-gradient $\Delta_i^{(s_i)}$, outer momentum $m_t$, arrival weight $\rho_t$;\\
alignment threshold $c_{\mathrm{ok}}$, correction constants $k_s, k_d, \kappa, \beta_{\max}, \varepsilon$
\end{tabular}
\AlgOutput
\begin{tabular}[t]{@{}l@{}}
Corrected update $\widehat{G}_t$
\end{tabular}
\State $\widehat{\Delta} \gets \Delta_i^{(s_i)}$
\For{each tensor block $b$}
    \State $u \gets (\Delta_i^{(s_i)})_b$;\quad $v \gets (m_t)_b$
    \If{$\|u\| < \varepsilon$ \textbf{or} $\|v\| < \varepsilon$}
        \State $\widehat{\Delta}_b \gets u$
        \Comment{Skip: direction unreliable}
    \Else
        \State $\hat{u} \gets u/\|u\|$;\quad $\hat{v} \gets v/\|v\|$
        \State $c_b \gets \hat{u}^\top \hat{v}$
        \Comment{Cosine alignment with momentum}
        \State $\mathrm{conf}_b \gets \|u\|/(\|u\| + \kappa\|v\| + \varepsilon)$
        \If{$c_b \geq c_{\mathrm{ok}}$}
            \State $\widehat{\Delta}_b \gets u$
            \Comment{Keep well-aligned blocks}
        \ElsIf{$c_b < 0$}
            \State $\beta_b \gets \min\{k_s(-c_b)\,\mathrm{conf}_b,\, \beta_{\max}\}$
            \State $\widehat{\Delta}_b \gets u - \beta_b c_b \|u\| \hat{v}$
            \Comment{Shrink the anti-momentum component}
        \Else
            \State $\lambda_b \gets \min\{k_d(1-c_b)\,\mathrm{conf}_b,\, 1\}$
            \State $\tilde{u} \gets (1-\lambda_b)\hat{u} + \lambda_b \hat{v}$
            \State $\widehat{\Delta}_b \gets \|u\|\,\tilde{u}/\max\{\|\tilde{u}\|, \varepsilon\}$
            \Comment{Weakly aligned: rotate toward momentum}
        \EndIf
    \EndIf
\EndFor
\State $\widehat{G}_t \gets \rho_t\,\widehat{\Delta}$
\State \Return $\widehat{G}_t$
\end{algorithmic}
\end{algorithm}

This appendix gives the procedural form of HeLoCo. The pseudocode shows the
two actions performed by the synchronizer during asynchronous training:
initializing an available worker with the look-ahead outer model, and updating
the outer model when a pseudo-gradient returns. Because workers run
independently, their updates need not arrive in the same order in which they
were initialized. The synchronizer therefore records the initialization step
$s_i$ for each worker and computes the staleness
$\tau_i(t)=t-s_i$ when the update is processed.





\subsection{Implementation and Compute Details}
\label{app:repro_compute}

All experiments were implemented in Python using \texttt{PyTorch 2.10} and \texttt{CUDA 12.6}. We use a TinyGPT-style decoder-only Transformer with 15M parameters and the GPT-2 tokenizer. The main software dependencies are \texttt{Transformers}, \texttt{NumPy}, \texttt{Pandas}, and \texttt{Matplotlib}. Since different experiment groups were run on different GPU nodes, we report the hardware configuration separately for each group.

For the experiments in Section~\ref{subsec:convergence_results}, each run uses
five workers. Each worker performs \texttt{20} local steps before returning a pseudo-gradient to the server, and training runs for \texttt{100} outer updates.
These experiments were run on a node with \texttt{4} NVIDIA A100 GPUs, \texttt{128} GB memory, and \texttt{8} CPU cores. The five workers were run in parallel, with each complete run taking approximately \texttt{60} minutes.

For the remaining results, we use a longer training setting with \texttt{80} local steps per worker update and \texttt{300} outer updates.  To study different levels of system heterogeneity, we vary the worker paces, which directly changes the amount of staleness in the asynchronous update stream. These experiments
were run on nodes with \texttt{8} NVIDIA H100 GPUs (\texttt{80} GB memory each), \texttt{112} CPU cores and \texttt{2} TB RAM. A training run with 24k steps requires approximately 15 minutes of runtime for one experiment, while using 5 GPUs - one for each worker. Within each experimental group, all methods use the same model, data preprocessing, optimizer settings, and evaluation protocol. To save compute time, a full model evaluation on the evaluation dataset was omitted after each outer update during the middle of each training run. This was only performed at the beginning and at the end of the training runs, which is why gaps appear in the middle of the loss plots.

\subsection{Training Hyperparameters}
The hyperparameters used for all experiments in Sections \ref{sec:noniidinsights} and \ref{app:deepiid} are summarized in table \ref{tab:hyperparameters}.\label{app:hyperparameters}
\begin{table}[h]
\centering
\small
\caption{Run configuration for non-iid experiments. There are different possibilities to manipulate weights. \textit{base} means the weighting factor is calculated as $\nicefrac{\sqrt{k}}{k}$, where k is the number of workers. This calculation was taken from the original async DiLoCo paper and showed the best results in our case. \textit{average} means a factor of $\nicefrac{1}{k}$ is used. For the inner learning rate the identical cosine scheduler from \citep{liu2024asynchronous} is used. Additional HeLoCo-specific parameters are: $k_{\text{d}}=1.0$, $c_{\text{ok}}=0.2$, , $\kappa=3.0$, $\beta_{\max}=0.5$, $k_{\text{s}}=0.5$}
\begin{tabular}{llll}

\toprule
Method & Weight Factor & Outer learning rate & Momentum \\
\midrule
async-HeLoCo & base & 0.7 & 0.9 \\
async-mla & base & 0.7 & 0.9 \\
async-nesterov & base & 0.07 & 0.9 \\
sync-nesterov & average & 0.7 & 0.9 \\
\bottomrule
\end{tabular}
\label{tab:hyperparameters}
\end{table}

\subsection{Deeper Insights into Non-IID Training}
\label{app:deepiid}

\textbf{Fixed shard-to-worker setup}

In Section \ref{sec:noniidinsights}, we observed that HeLoCo achieves a lower overall loss than the other algorithms in the non-IID setting where shards are permanently assigned to workers. This conclusion was based on the average loss across all languages. In the following, we analyze in more detail how different levels of staleness affect the representation of individual shards (i.e., languages) within the global model.

\begin{figure}[t]
    \centering

    \begin{subfigure}[b]{\linewidth}
        \centering
        \includegraphics[width=\linewidth]{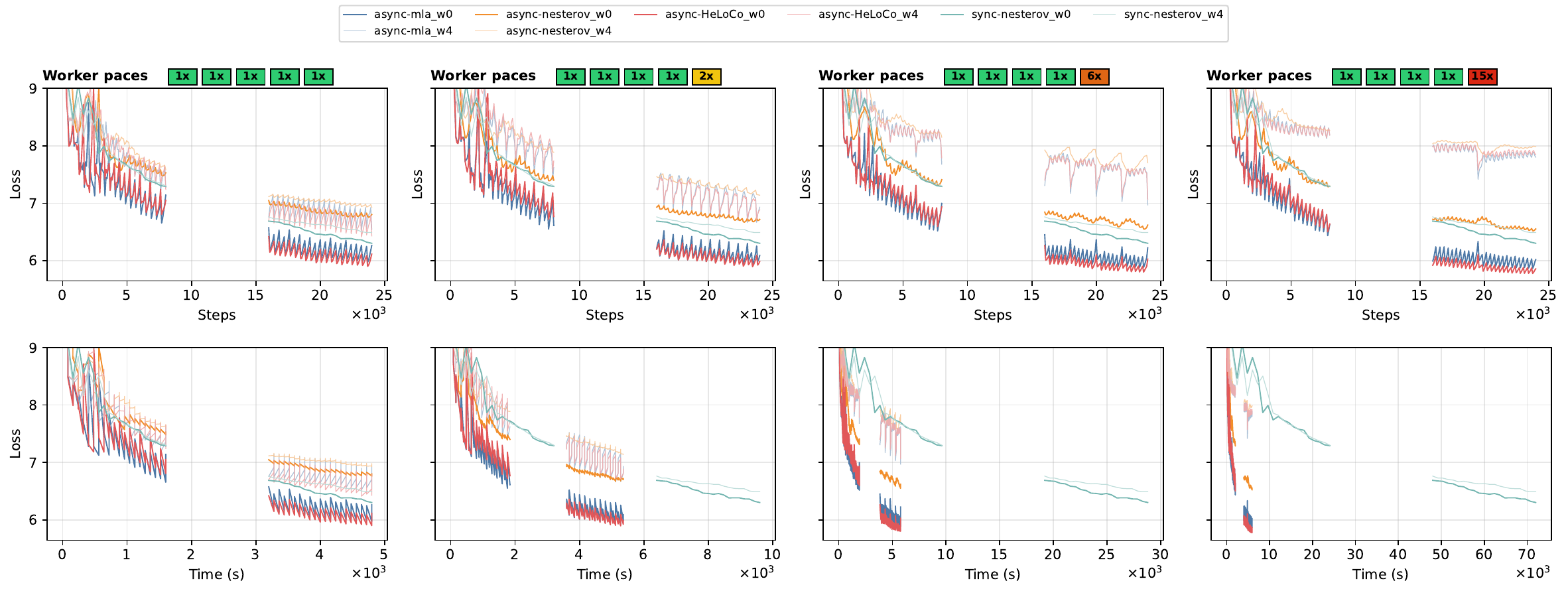}

    \end{subfigure}
    \begin{subfigure}[b]{\linewidth}
        \centering
        \includegraphics[width=\linewidth]{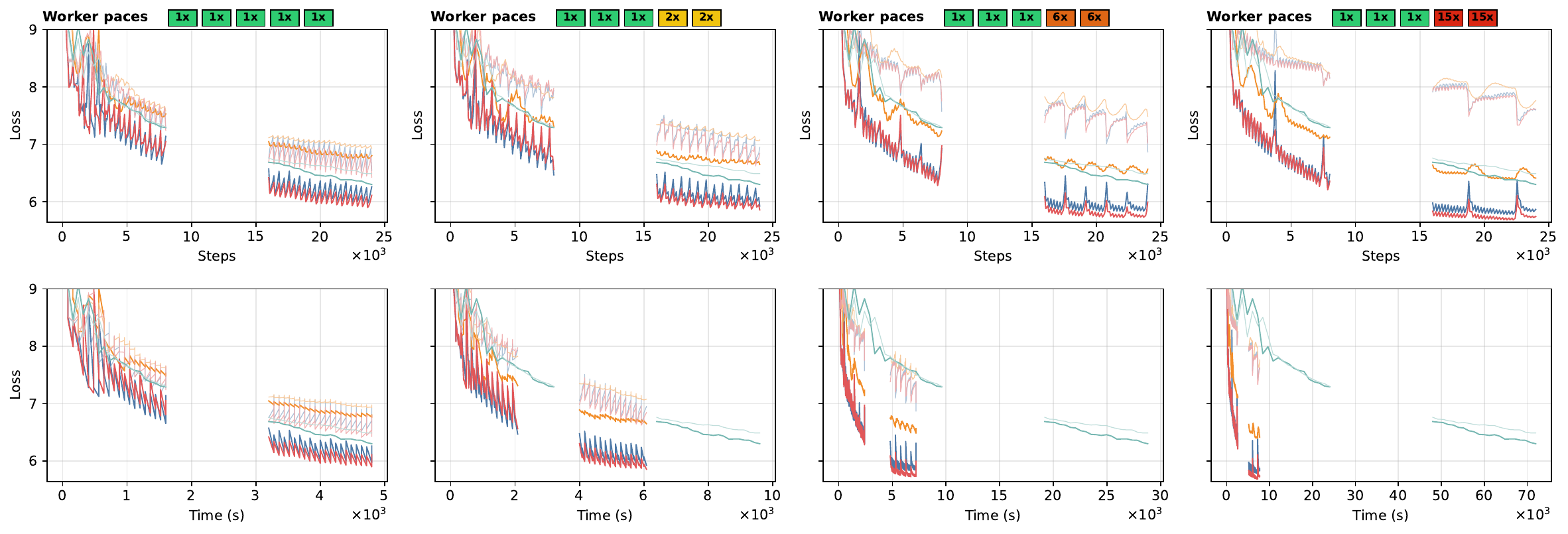}

    \end{subfigure}
     \begin{subfigure}[b]{\linewidth}
        \centering
        \includegraphics[width=\linewidth]{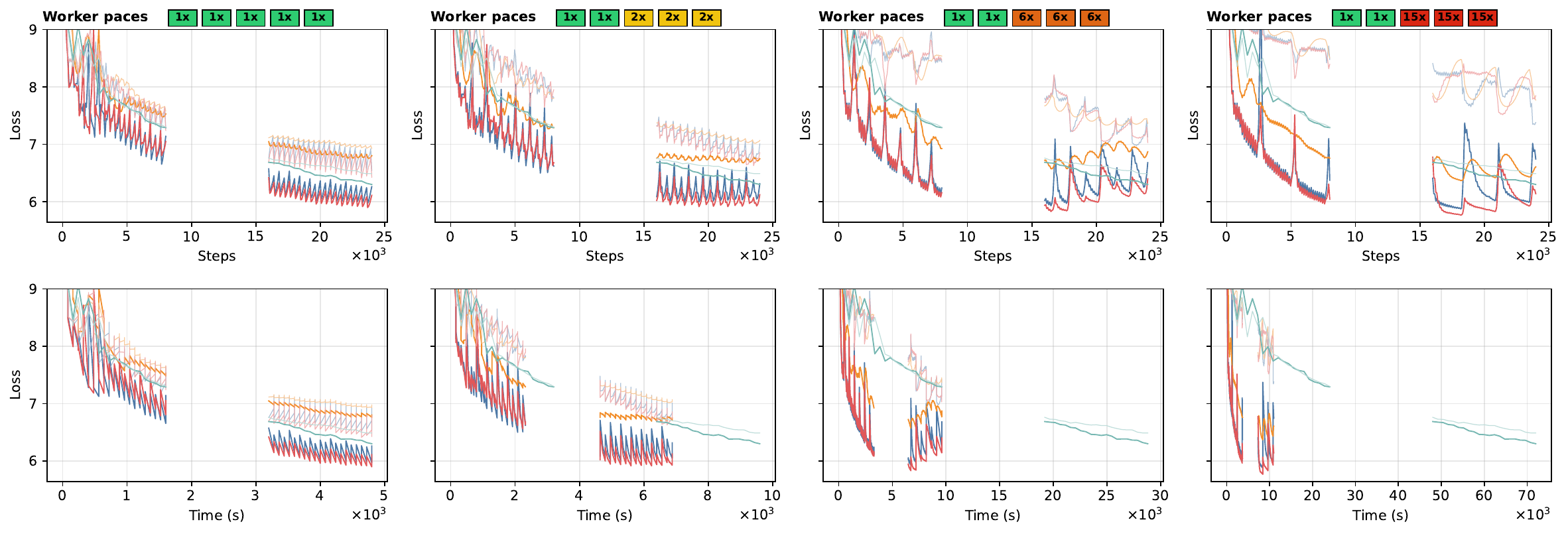}
 
    \end{subfigure}
     \begin{subfigure}[b]{\linewidth}
        \centering
        \includegraphics[width=\linewidth]{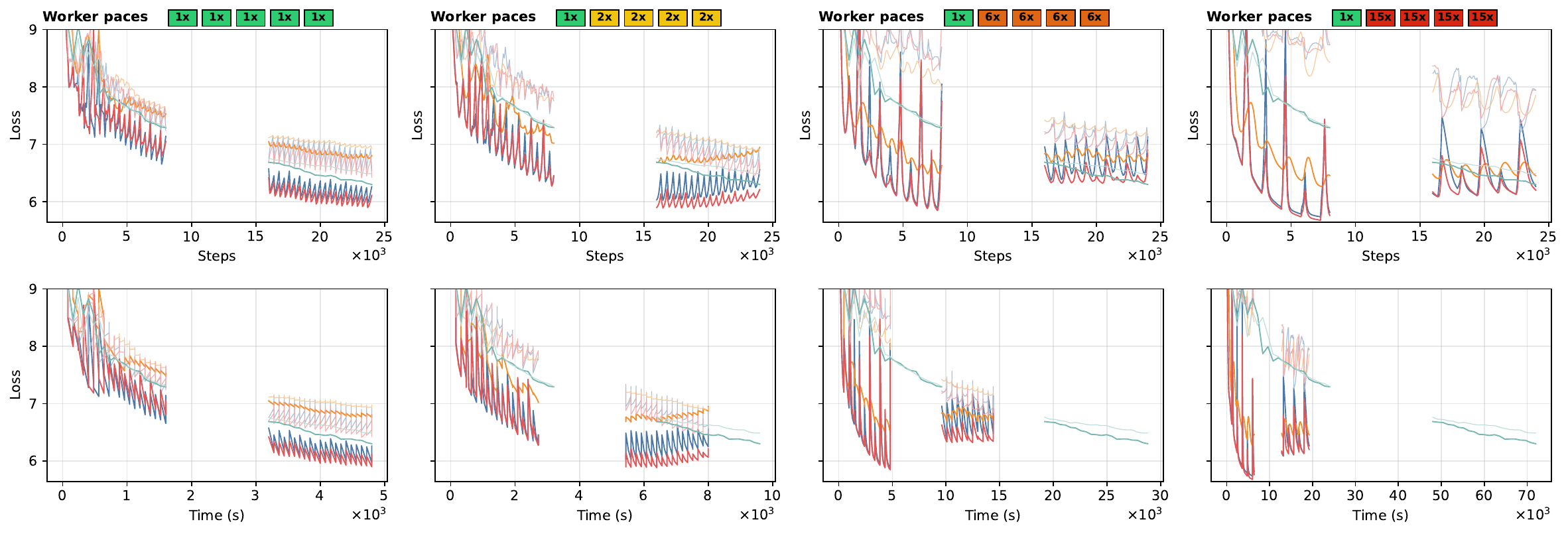}
      
    \end{subfigure}
    \caption{Influence of the number and degree of stale workers on the validation loss for different languages in the server model (fixed setup). In each plot, worker 0 (lowest staleness) and worker 4 (highest staleness) are shown.}
    \label{fig:combined_diffpacelarge}
\end{figure}

In Figure \ref{fig:combined_diffpacelarge}, we first analyze the behavior when introducing a single stale worker, starting from the homogeneous case and examining what happens when the staleness of one worker is increased.
 When a worker with pace 2 (seconds/iteration) is added, the loss of the fast worker (pace 1 second/iteration) changes only marginally, while the loss associated with worker 4 (the stale worker) increases immediately. As the staleness is increased further, this effect becomes more pronounced. The loss associated with the stale worker continues to deteriorate, and its corresponding language becomes increasingly underrepresented in the global model. This behavior remains stable for up to three stale workers with pace 2.

Once the number of stale workers reaches four, or when three workers exhibit a staleness of 6 or higher, a different behavior emerges for both MLA and HeLoCo. The loss associated with the fast worker (worker 0) begins to increase again during training, while the losses of the slow workers continue to decrease. This suggests that beyond a certain level of staleness, the global model starts to forget information associated with the fast workers language in favor of the stale workers. In other words, the model appears to lose previously acquired knowledge for well-represented shards while adapting to underrepresented ones. 

\begin{figure}[t]
    \centering

    \begin{subfigure}[b]{\linewidth}
        \centering
        \includegraphics[width=\linewidth]{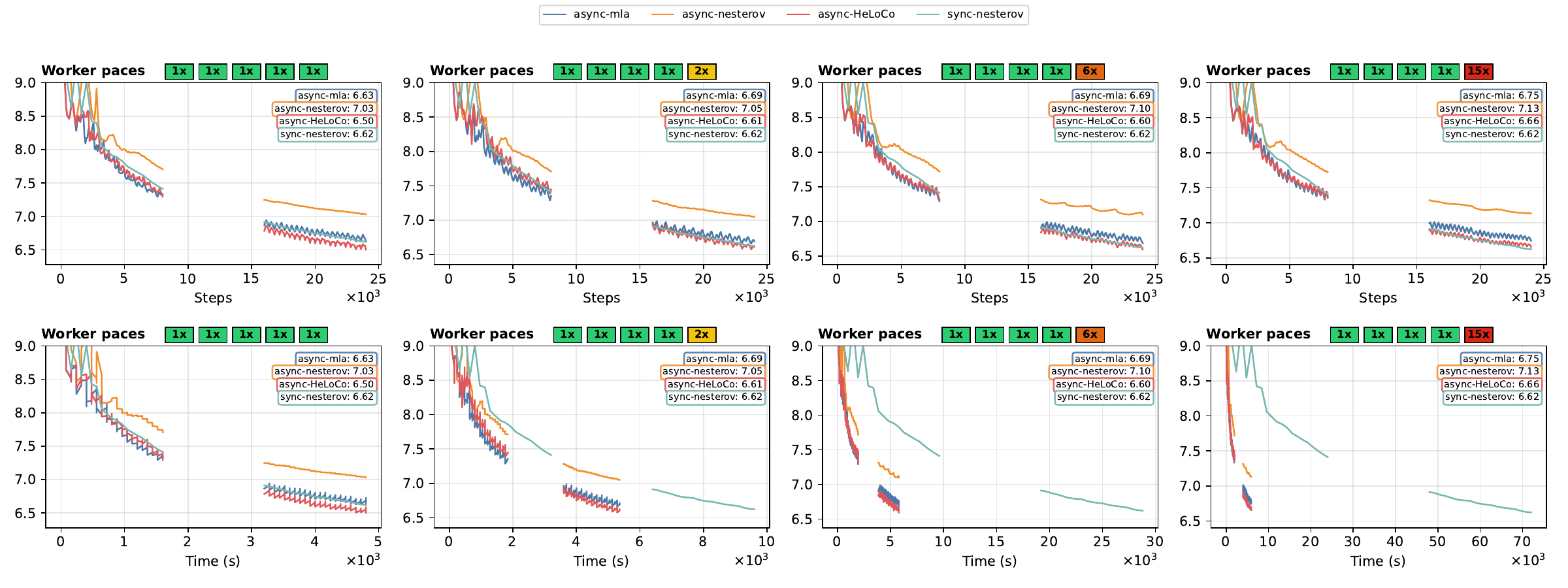}
        
    \end{subfigure}
    \begin{subfigure}[b]{\linewidth}
        \centering
        \includegraphics[width=\linewidth]{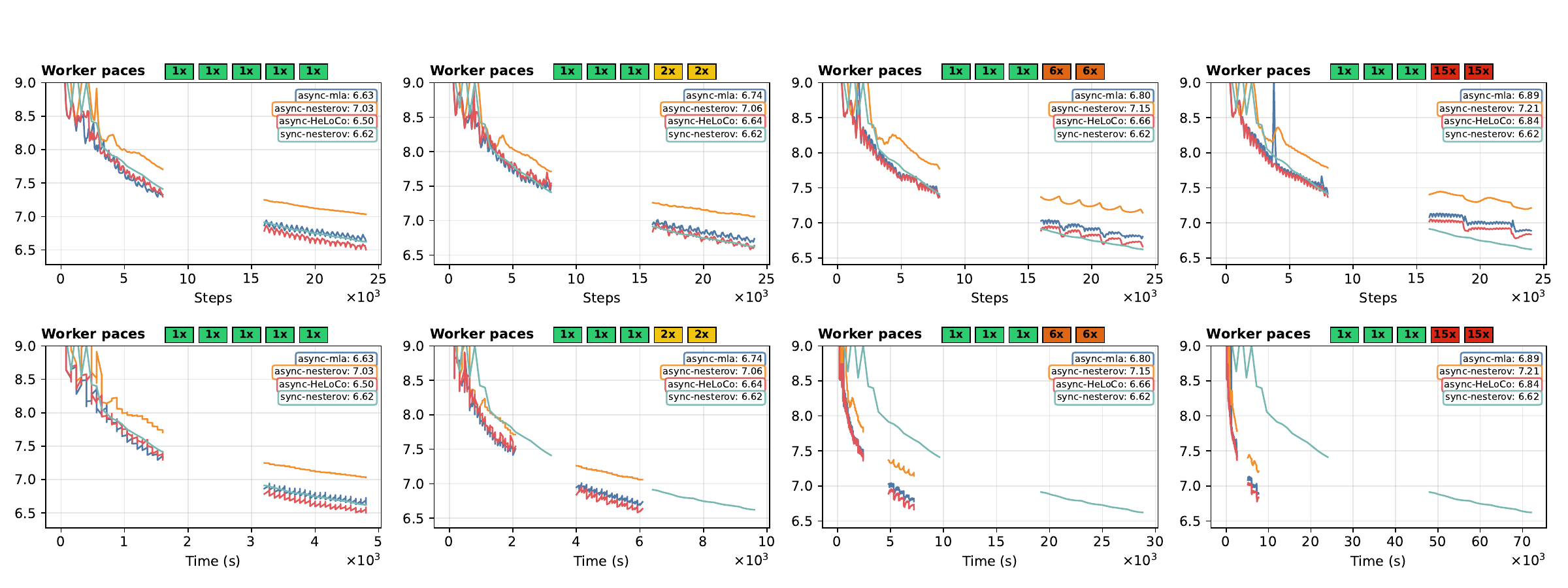}
    
    \end{subfigure}
     \begin{subfigure}[b]{\linewidth}
        \centering
        \includegraphics[width=\linewidth]{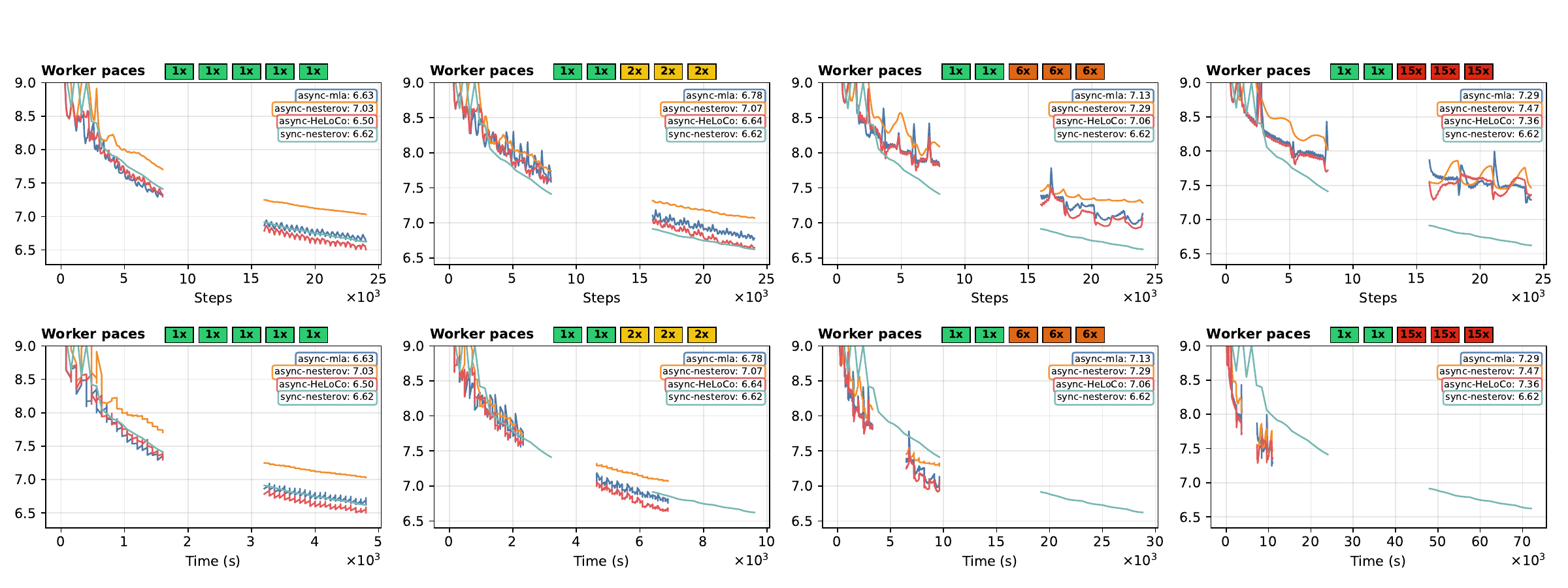}
        
    \end{subfigure}
     \begin{subfigure}[b]{\linewidth}
        \centering
        \includegraphics[width=\linewidth]{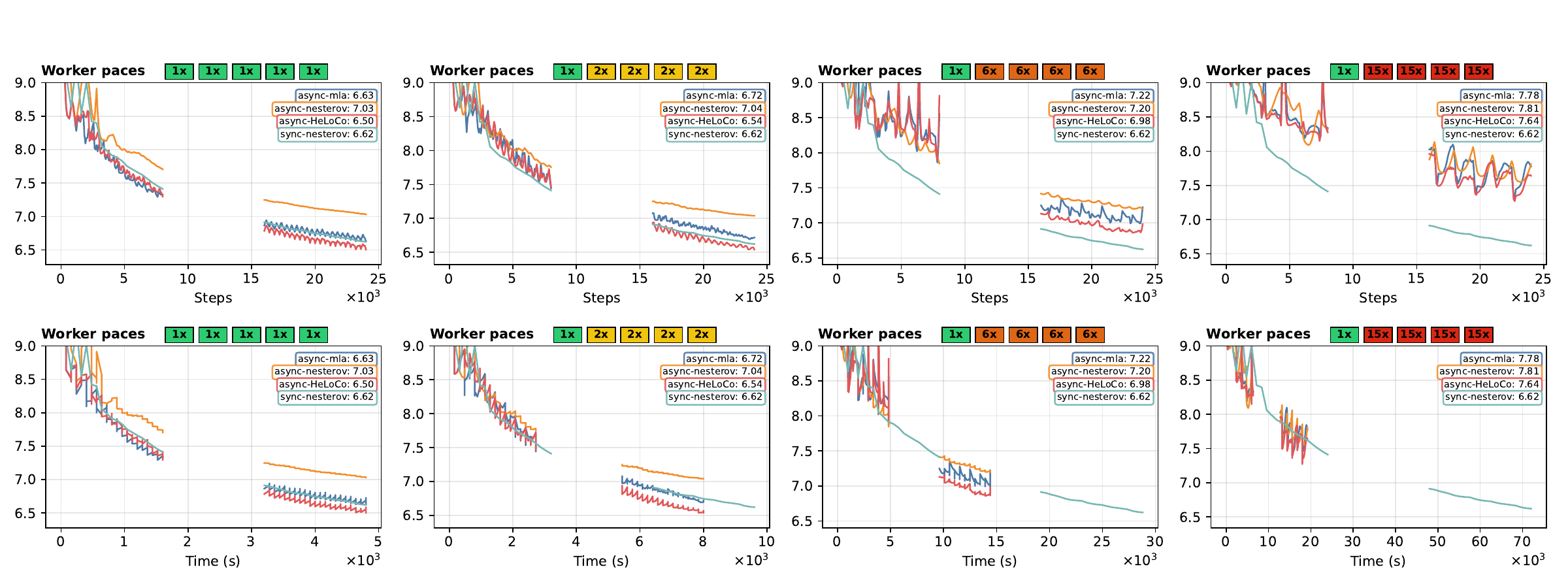}
        
    \end{subfigure}
    \caption{Influence of the number and degree of stale workers on the mean validation loss of the server model.}
    \label{fig:combined_diffpacelargem}
\end{figure}

Interestingly, this effect is not visible when considering only the mean validation loss across all languages, as shown in Figure \ref{fig:combined_diffpacelargem}. The degradation in performance for worker 0 is compensated by improvements for the stale workers, resulting in a relatively stable average loss. For future work, it may be interesting to investigate this behavior in more detail to better understand under which conditions this forgetting effect occurs and how it might be mitigated through appropriate training strategies.

\textbf{Flexible shard-to-worker setup}

In this section, we investigate a flexible shard-to-worker assignment strategy, in which each worker dynamically selects the shard that has been processed the least overall. This setup corresponds to the scenario used in the original async DiLoCo work of \citep{liu2024asynchronous}. The key difference in our experiments is that the shards remain highly non-IID.

\begin{figure}[t]
    \centering

    \begin{subfigure}[b]{\linewidth}
        \centering
        \includegraphics[width=\linewidth]{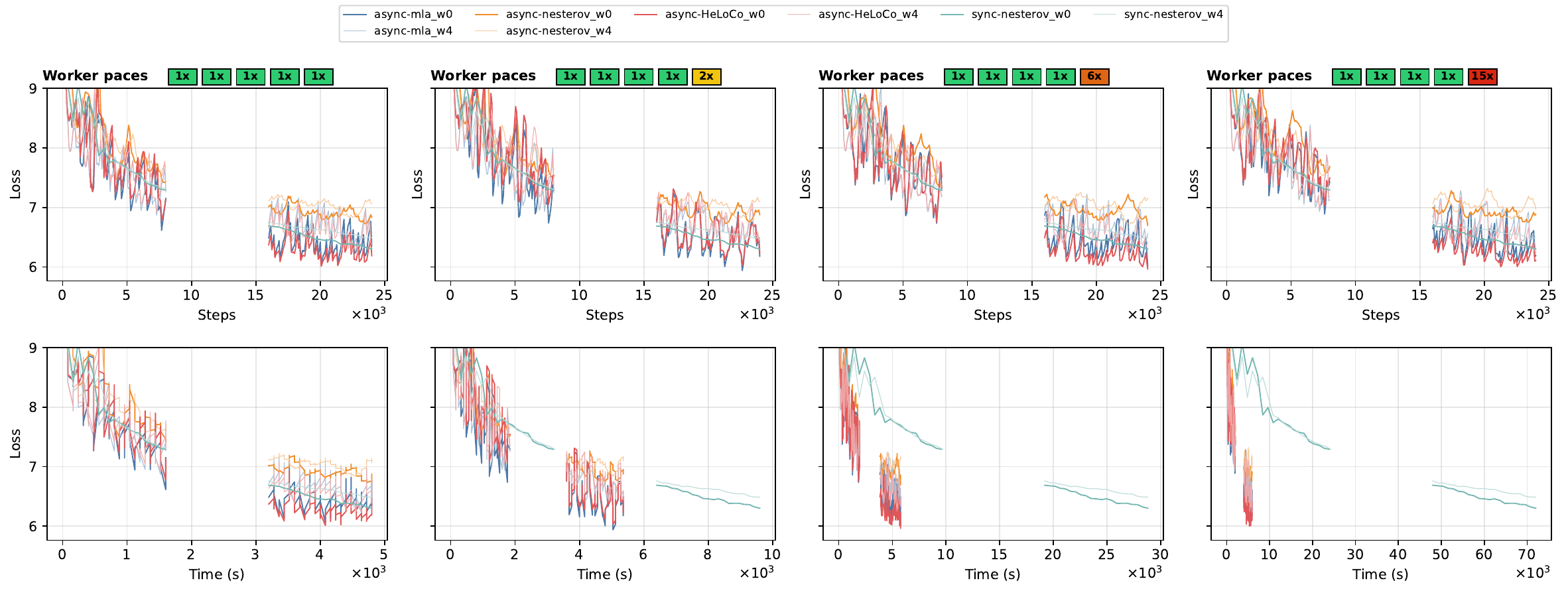}
      
    \end{subfigure}
    \begin{subfigure}[b]{\linewidth}
        \centering
        \includegraphics[width=\linewidth]{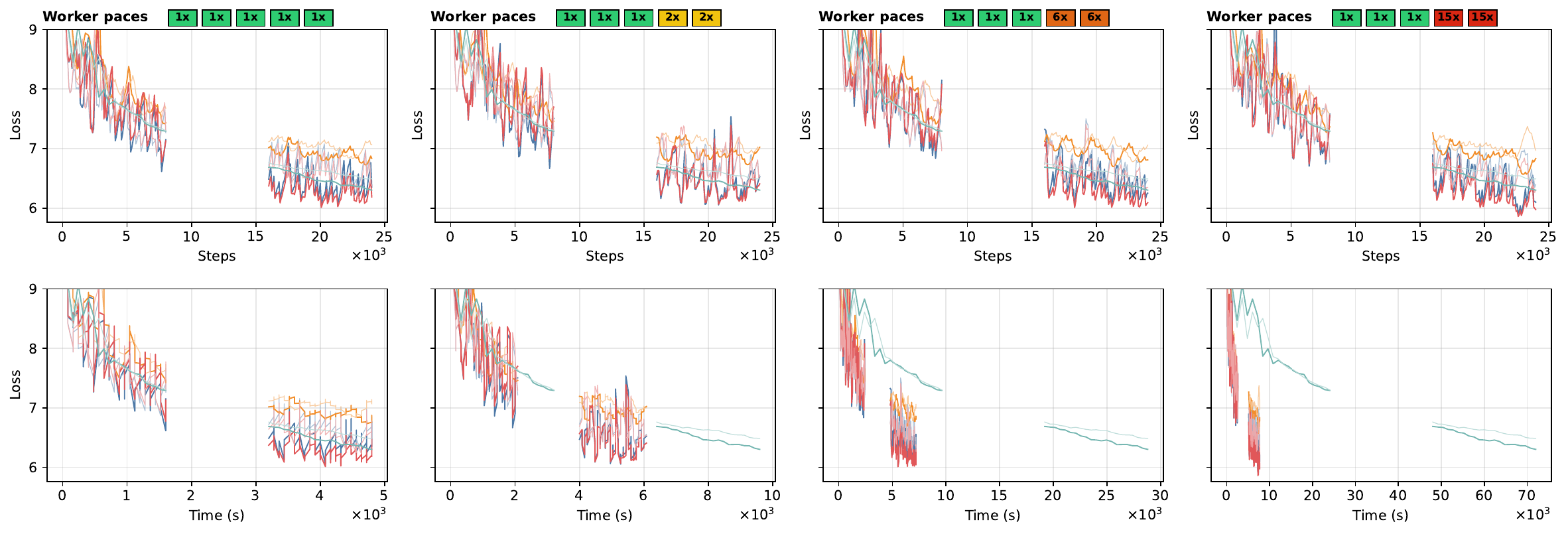}
       
    \end{subfigure}
     \begin{subfigure}[b]{\linewidth}
        \centering
        \includegraphics[width=\linewidth]{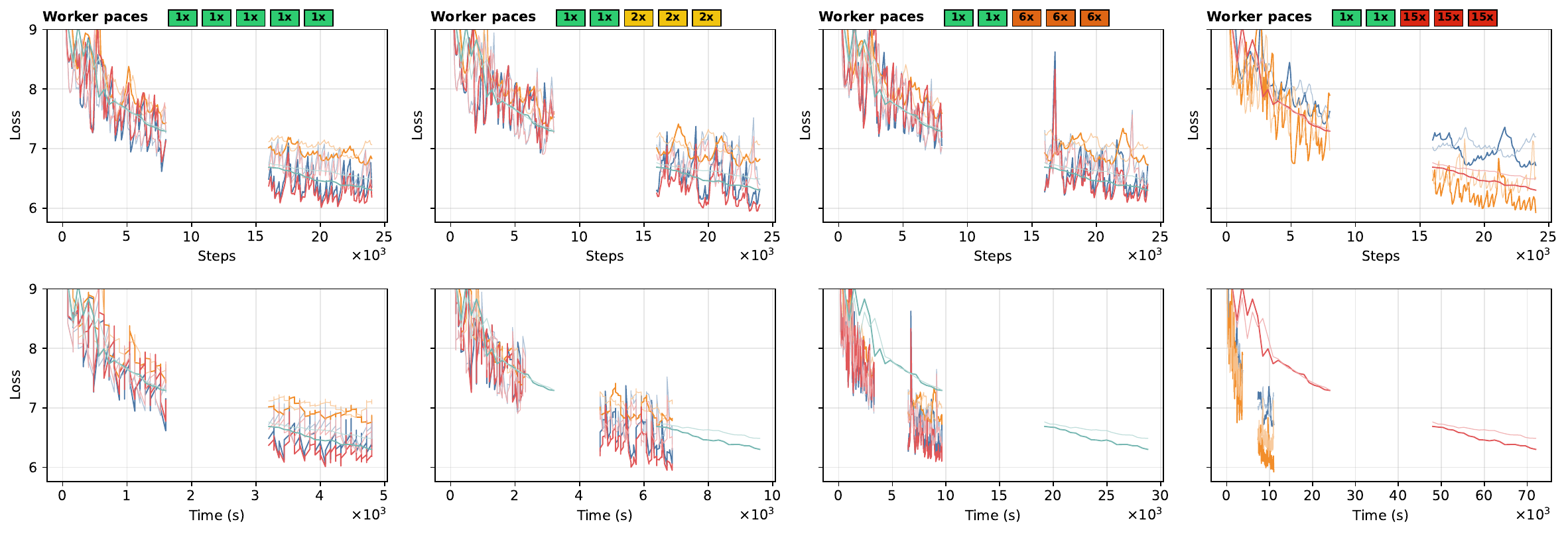}
        
    \end{subfigure}
     \begin{subfigure}[b]{\linewidth}
        \centering
        \includegraphics[width=\linewidth]{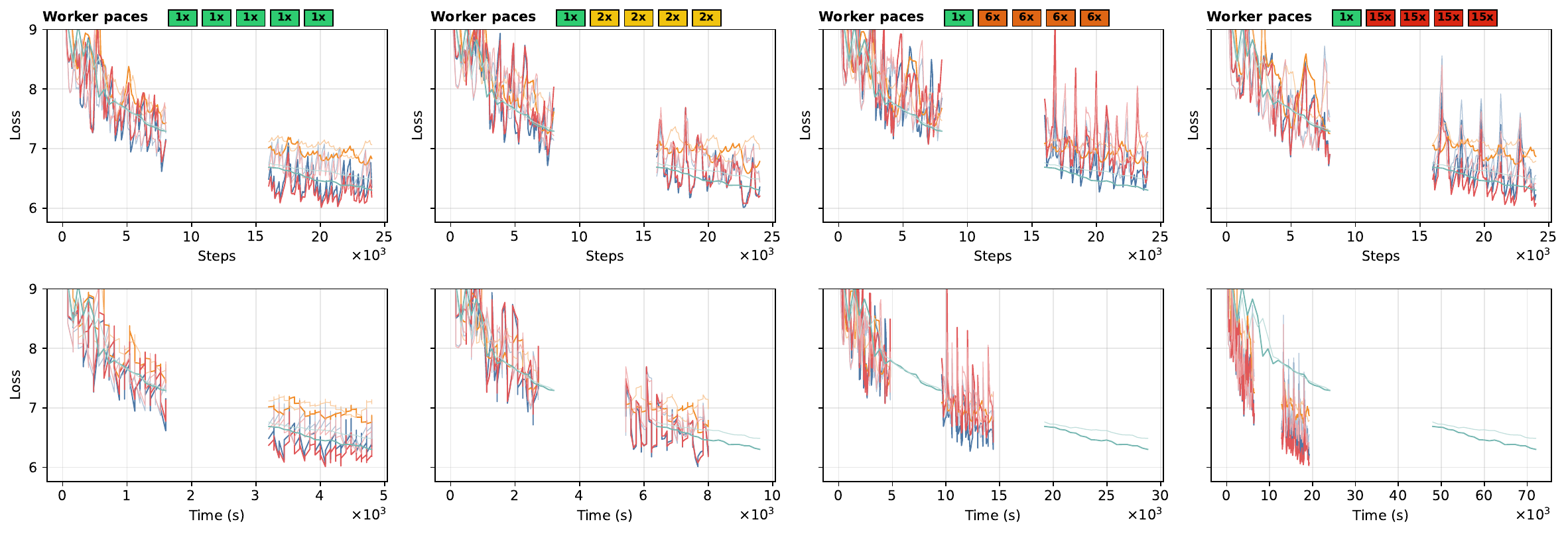}
       
    \end{subfigure}
    \caption{Influence of the number and degree of stale workers on the validation loss for different languages in the server model (flexible setup). In each plot, worker 0 (lowest staleness) and worker 4 (highest staleness) are shown.}
    \label{fig:combined_diffpacelargef}
\end{figure}

Compared to the fixed shard-to-worker setup, Figure \ref{fig:combined_diffpacelargef} shows that the discrepancy between fast and slow workers has a much smaller impact on the language-specific losses. The flexible assignment strategy leads to a more balanced representation of the different shards, because fast workers are no longer restricted to a single shard and can contribute to multiple languages over time.

Finally, we investigate whether discarding stale gradients can improve the overall training performance. This means for example, that in the case $(1,1,6,6,6)$ the pseudo-gradient of three stale workers is set to 0 in order to suppress their influence on the server model.

\begin{figure}[t]
    \centering

    \begin{subfigure}[b]{\linewidth}
        \centering
        \includegraphics[width=\linewidth]{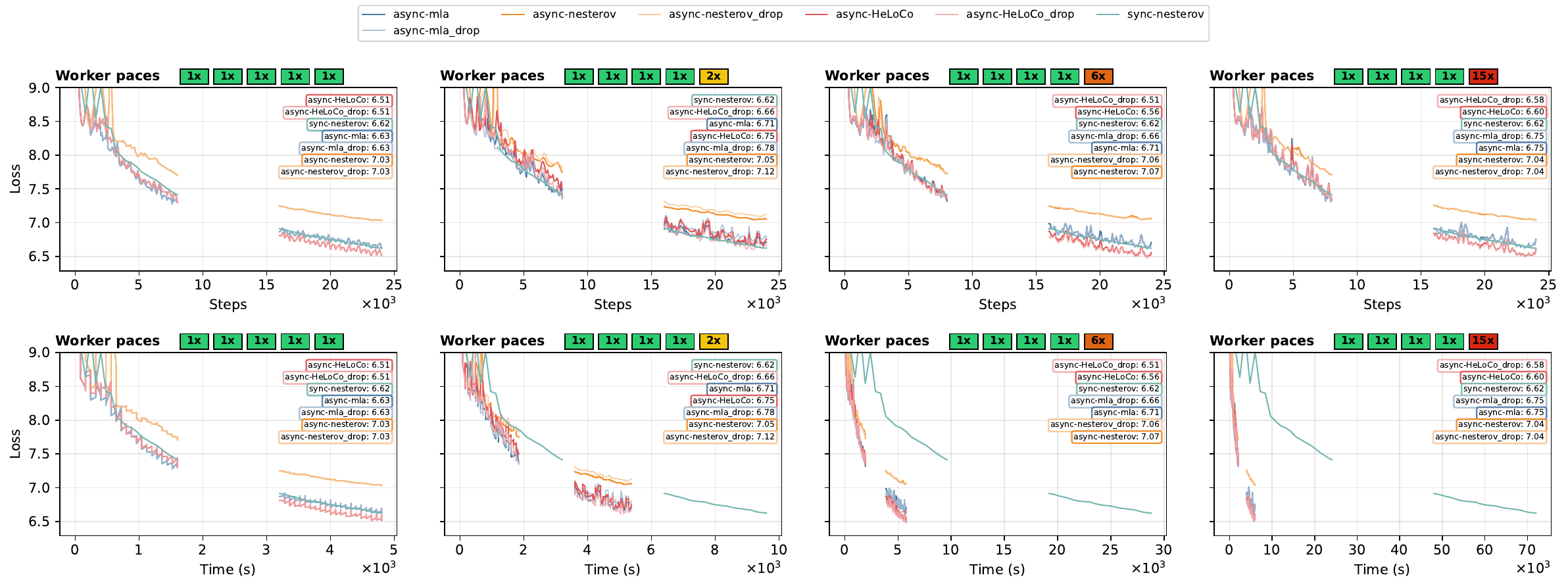}
      
    \end{subfigure}
    \begin{subfigure}[b]{\linewidth}
        \centering
        \includegraphics[width=\linewidth]{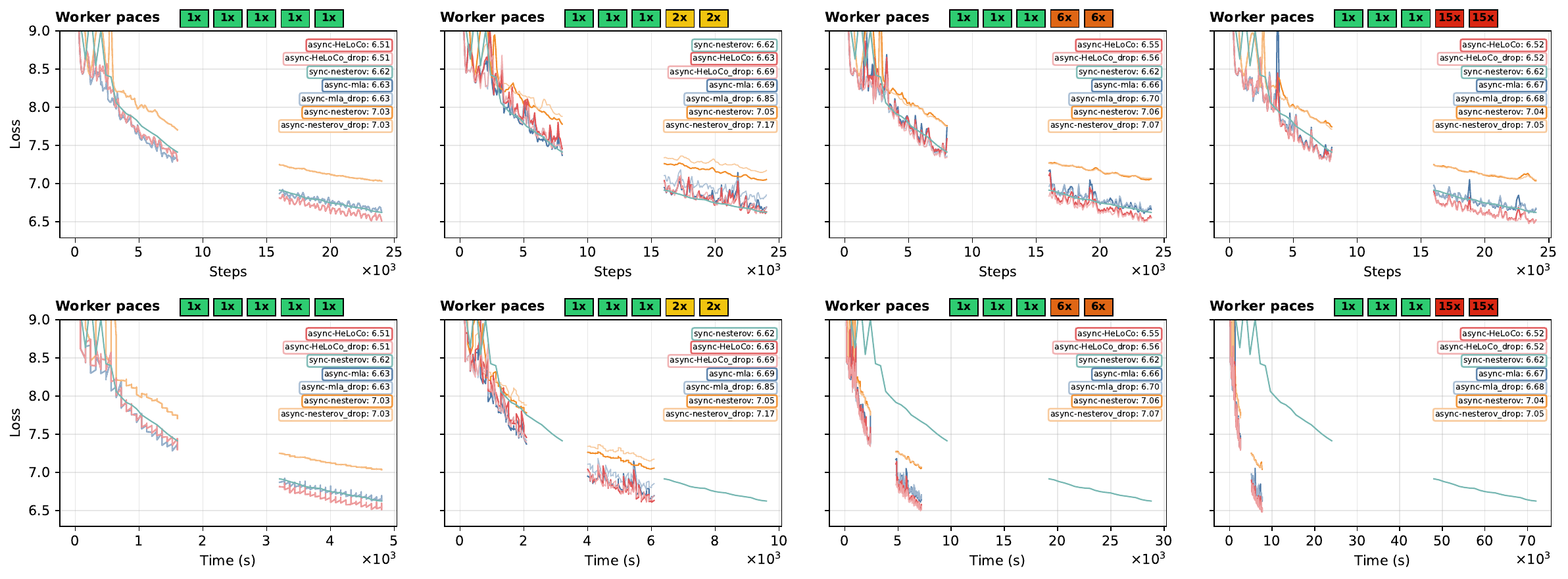}
        
    \end{subfigure}
     \begin{subfigure}[b]{\linewidth}
        \centering
        \includegraphics[width=\linewidth]{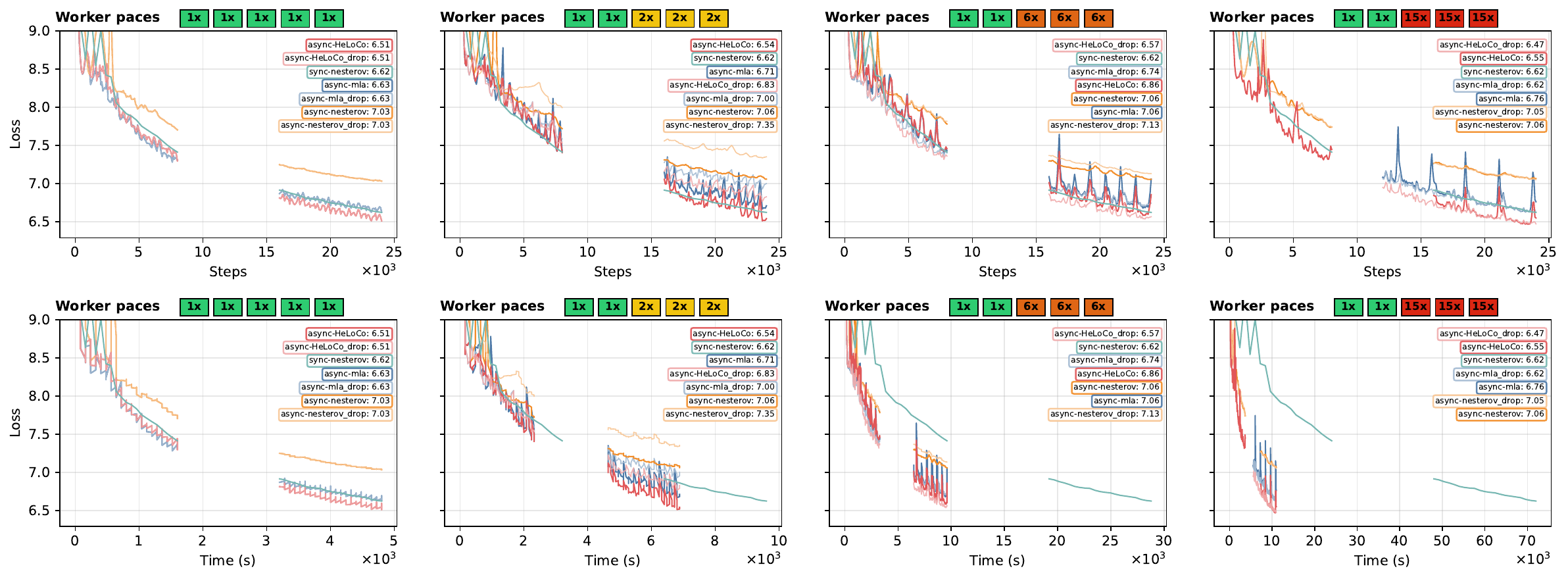}
        
    \end{subfigure}
     \begin{subfigure}[b]{\linewidth}
        \centering
        \includegraphics[width=\linewidth]{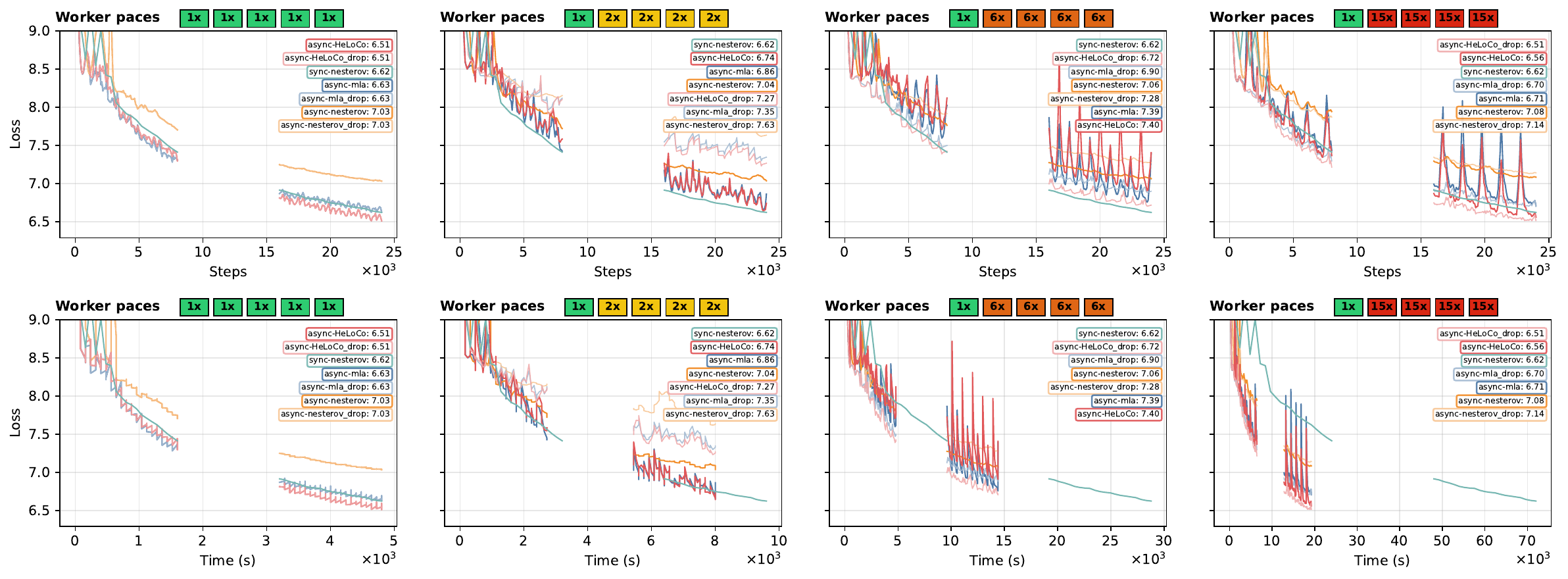}
       
    \end{subfigure}
    \caption{Influence of dropping pseudo-gradients from stale workers on the mean validation loss over all languages.}
    \label{fig:combined_diffpacelargefd}
\end{figure}

Figure \ref{fig:combined_diffpacelargefd} shows that discarding gradients in scenarios with more than three stale workers and a staleness of 6 or greater consistently improves the performance of MLA-based algorithms, while it leads to degraded loss for asynchronous Nesterov. This suggests that neither MLA-based HeLoCo nor the pure MLA approach is able to effectively exploit highly stale updates in this setting, in contrast to asynchronous Nesterov. Rather than contributing useful information to training, these stale updates appear to impair the quality of the global model. Overall, this behavior indicates that there is still room for improvement, and that developing an adaptive mechanism to determine which workers provide beneficial updates and which do not could be a promising direction for future work.

\clearpage

\end{document}